\documentclass{emulateapj}

\usepackage{lscape}
\usepackage{apjfonts}
\usepackage{graphicx}
\usepackage{natbib}
\slugcomment{Received 2010 March 10; accepted 2010 April 7; to appear in ApJL}

\newcommand{\angstrom}{{\rm \AA}}

\newcommand{\hbeta}{H{$\beta$}}
\newcommand{\halpha}{H{$\alpha$}}

\newcommand{\OIII}{[O{\sevenrm\,III}]}

\newcommand{\OIIIa}{[O{\sevenrm\,III}]\,$\lambda$4959}
\newcommand{\OIIIb}{[O{\sevenrm\,III}]\,$\lambda$5007}
\newcommand{\OIIIc}{[O{\sevenrm\,III}]\,$\lambda\lambda$4959,5007}
\newcommand{\NII}{[N{\sevenrm\,II}]}
\newcommand{\NIIb}{[N{\sevenrm\,II}]\,$\lambda$6584}

\newcommand{\SIIa}{[S{\sevenrm\,II}]\,$\lambda$6717}
\newcommand{\SIIb}{[S{\sevenrm\,II}]\,$\lambda$6731}

 \font\sevenrm=cmr7 scaled 1000

\begin{document}

\title{Discovery of Four kpc-Scale Binary AGNs\altaffilmark{1}}

\shorttitle{BINARY AGNs}

\shortauthors{LIU, GREENE, SHEN, \& STRAUSS}
\author{Xin Liu\altaffilmark{2}, Jenny E. Greene\altaffilmark{2,3}, Yue
Shen\altaffilmark{4}, and Michael A. Strauss\altaffilmark{2}}

\altaffiltext{1}{This paper includes data gathered with the 6.5
meter Magellan Telescopes located at Las Campanas Observatory,
Chile, and with the Apache Point Observatory 3.5-meter telescope,
which is owned and operated by the Astrophysical Research
Consortium.}

\altaffiltext{2}{Department of Astrophysical Sciences, Princeton
University, Peyton Hall -- Ivy Lane, Princeton, NJ 08544}

\altaffiltext{3}{Princeton-Carnegie Fellow}

\altaffiltext{4}{Harvard-Smithsonian Center for Astrophysics, 60
Garden St., MS-51, Cambridge, MA 02138}

\begin{abstract}
We report the discovery of four kpc-scale binary AGNs. These
objects were originally selected from the Sloan Digital Sky Survey
based on double-peaked \OIIIc\ emission lines in their fiber
spectra.  The double peaks could result from pairing active
supermassive black holes (SMBHs) in a galaxy merger, or could be
due to bulk motions of narrow-line region gas around a single
SMBH.  Deep near-infrared (NIR) images and optical slit spectra
obtained from the Magellan 6.5 m and the APO 3.5 m telescopes
strongly support the binary SMBH scenario for the four objects. In
each system, the NIR images reveal tidal features and double
stellar components with a projected separation of several kpc, while
optical slit spectra show two Seyfert 2 nuclei spatially
coincident with the stellar components, with line-of-sight velocity
offsets of a few hundred km s$^{-1}$.  These objects were drawn
from a sample of only 43 objects, demonstrating the efficiency of
this technique to find kpc-scale binary AGNs.
\end{abstract}

\keywords{black hole physics -- galaxies: active -- galaxies:
interactions -- galaxies: nuclei -- galaxies: Seyfert -- quasars:
general}

\section{Introduction}\label{sec:intro}

Despite decades of searching, and strong theoretical reasons to
believe they exist, merging active supermassive black holes
(SMBHs) remain difficult to find.  Here we present the first four
secure candidates from our ongoing search for kpc-scale
binary\footnote{We refer to both a bound pair of SMBHs in a
Keplerian orbit, and a pair of SMBHs at large separations where
the galactic potential dominates, as {\it binaries}.} AGNs.

It is widely appreciated that galaxies are built up hierarchically
via mergers \citep{toomre72}.  Since most massive galaxies are
believed to harbor a central SMBH \citep{kormendy95}, galaxy
mergers would result in the formation of binary SMBHs
\citep{begelman80,milosavljevic01}.  Major mergers between
galaxies are responsible, at least in some models, for triggering
black hole accretion activity in the most luminous AGNs.
Merger-based models do a decent job of fitting observational data
on local SMBH demographics and AGN statistics at different cosmic
epochs \citep[e.g.,][]{kauffmann00,volonteri03,
wyithe03,hopkins08,shen09}, and on the core properties of
elliptical galaxies \citep[e.g.,][]{faber97,kormendy09}, but
direct observational evidence for binary SMBHs remains
surprisingly scarce \citep[e.g.,][]{komossa03,bianchi08,rodriguez06,comerford09,green10}.

\citet{gerke07} and \citet{comerford08} reported two binary-AGN
candidates in the DEEP2 galaxy redshift survey \citep{davis03},
which show spatially resolved (on $\sim$ kpc scales) double-peaked
\OIII~$\lambda 5007$ emission with velocity offsets of a few
hundred ${\rm km\,s^{-1}}$.  We have selected a similar sample of
167 double-peaked \OIII\ narrow-line AGNs (see also
\citealt{smith09} and \citealt{wang09} for similar statistical
studies, as well as work on individual interesting objects, e.g.
\citealt{xu09}) from the Seventh Data Release of the Sloan Digital
Sky Survey \citep[SDSS;][]{york00,SDSSDR7} and studied their
host-galaxy and emission-line properties and selection
completeness \citep{liu10}. The double-peaked features could be
due either to narrow-line region (NLR) kinematics such as outflows
or rotating disks \citep[e.g.,][]{axon98,veilleux01,crenshaw09},
or to a pair of AGNs.  To discriminate between these scenarios for
individual objects and to identify bona fide AGN pairs, we are
conducting deep near-infrared (NIR) imaging and spatially resolved
optical spectroscopy for the objects in the double-peaked
narrow-line sample. Here we report initial results on the
discovery of four binary AGNs, SDSSJ$110851.04+065901.4$,
SDSSJ$113126.08-020459.2$, SDSSJ$114642.47+511029.6$, and
SDSSJ$133226.34+060627.4$.  Deep NIR imaging revealed double
stellar bulges in SDSSJ$110851.04+065901.4$,
SDSSJ$113126.08-020459.2$, and SDSSJ$133226.34+060627.4$, while
SDSSJ$114642.47+511029.6$'s SDSS image shows clear evidence of a 
recent merger.  We have obtained slit spectroscopy of all four sources.
We describe our observations data analysis in \S\S
\ref{subsec:panic}-\ref{subsec:ldss3}, followed by discussion of
the nature of the ionizing sources in \S \ref{subsec:result}. We
discuss our results and conclude in \S \ref{sec:discuss}.  A
cosmology with $\Omega_m = 0.3$, $\Omega_{\Lambda} = 0.7$, and $h
= 0.7$ is assumed throughout.

\section{Observations and Data Analysis}\label{sec:obs}

\subsection{NIR Imaging}\label{subsec:panic}

\begin{figure}
  \centering
    \includegraphics[width=40mm]{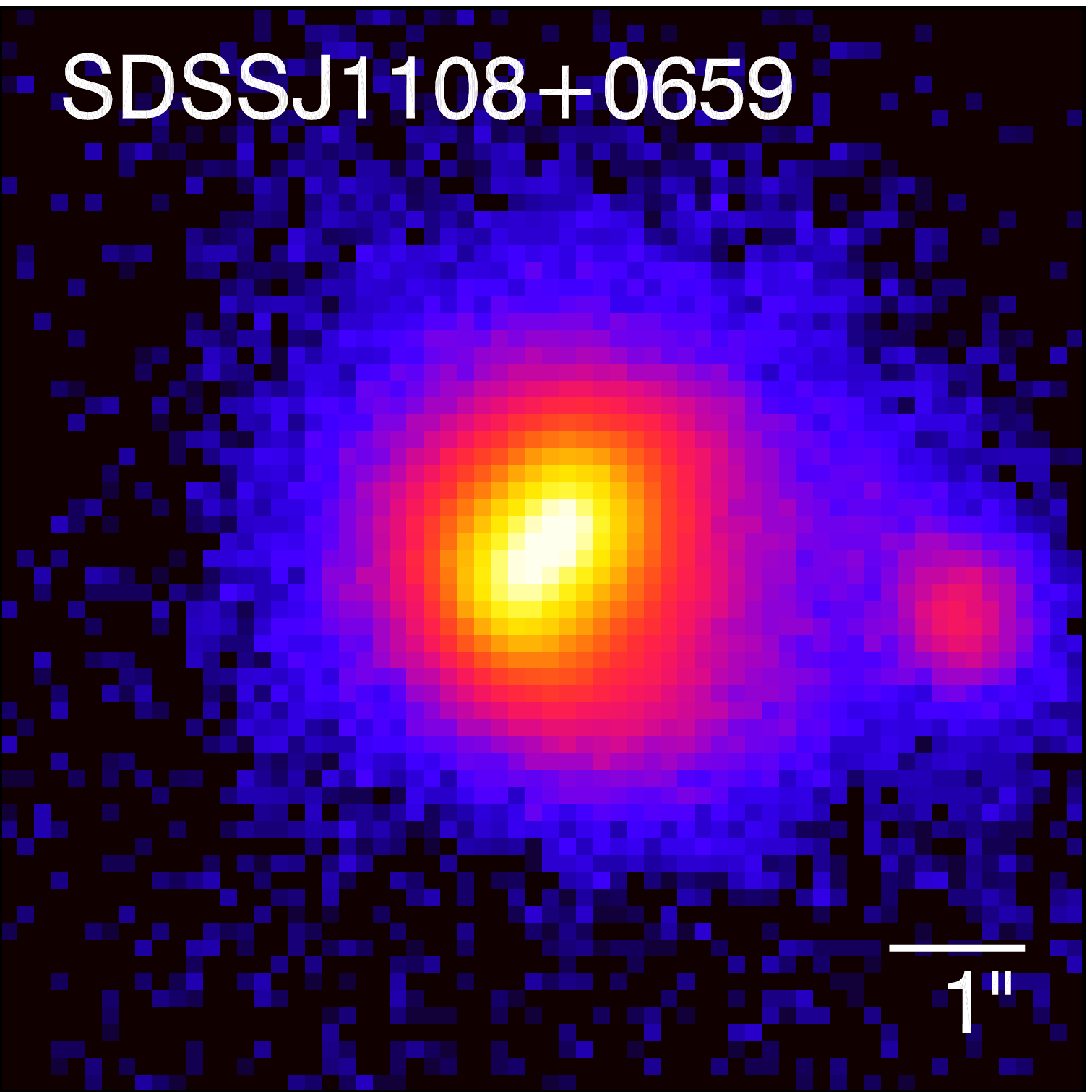}
    \includegraphics[width=40mm]{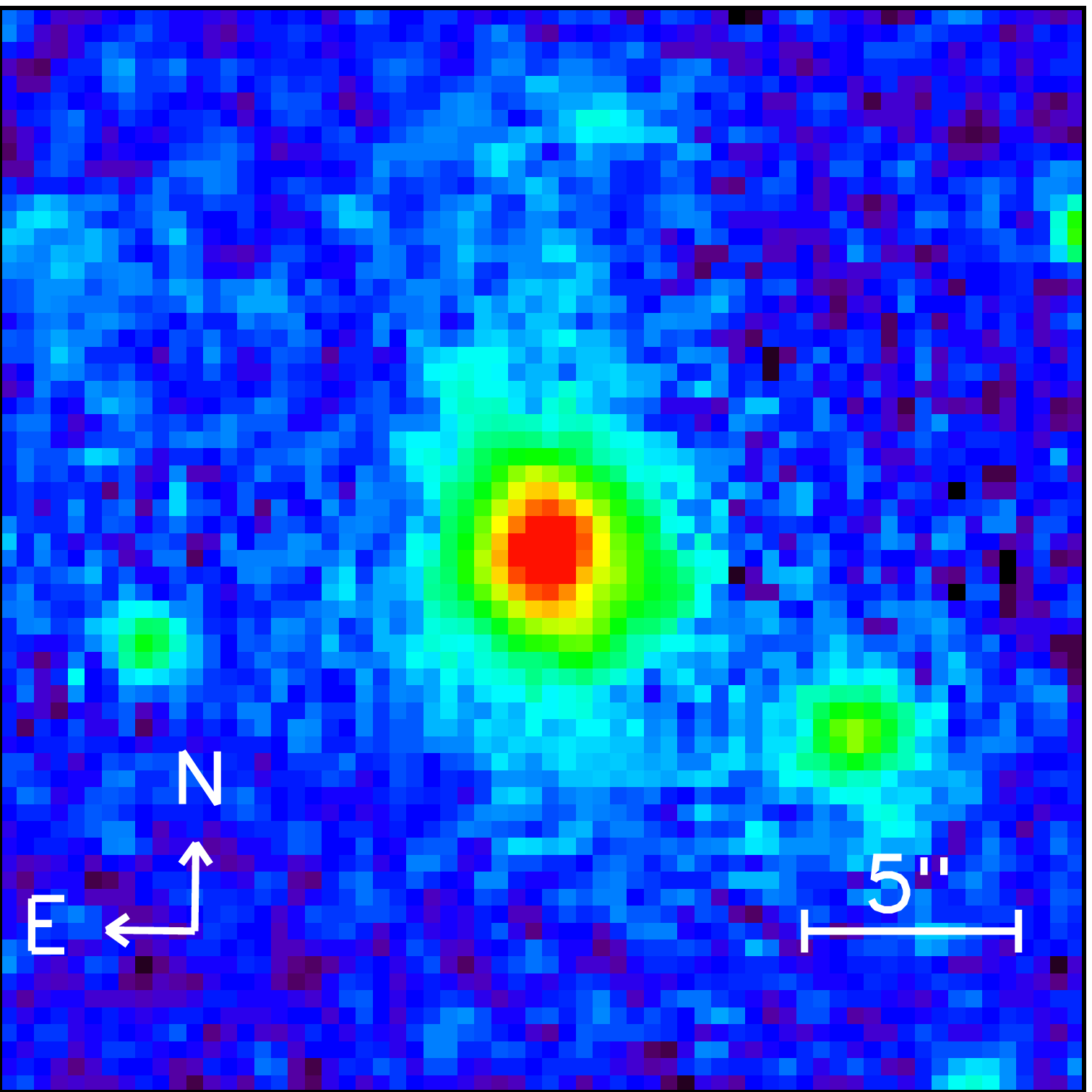}
    \includegraphics[width=40mm]{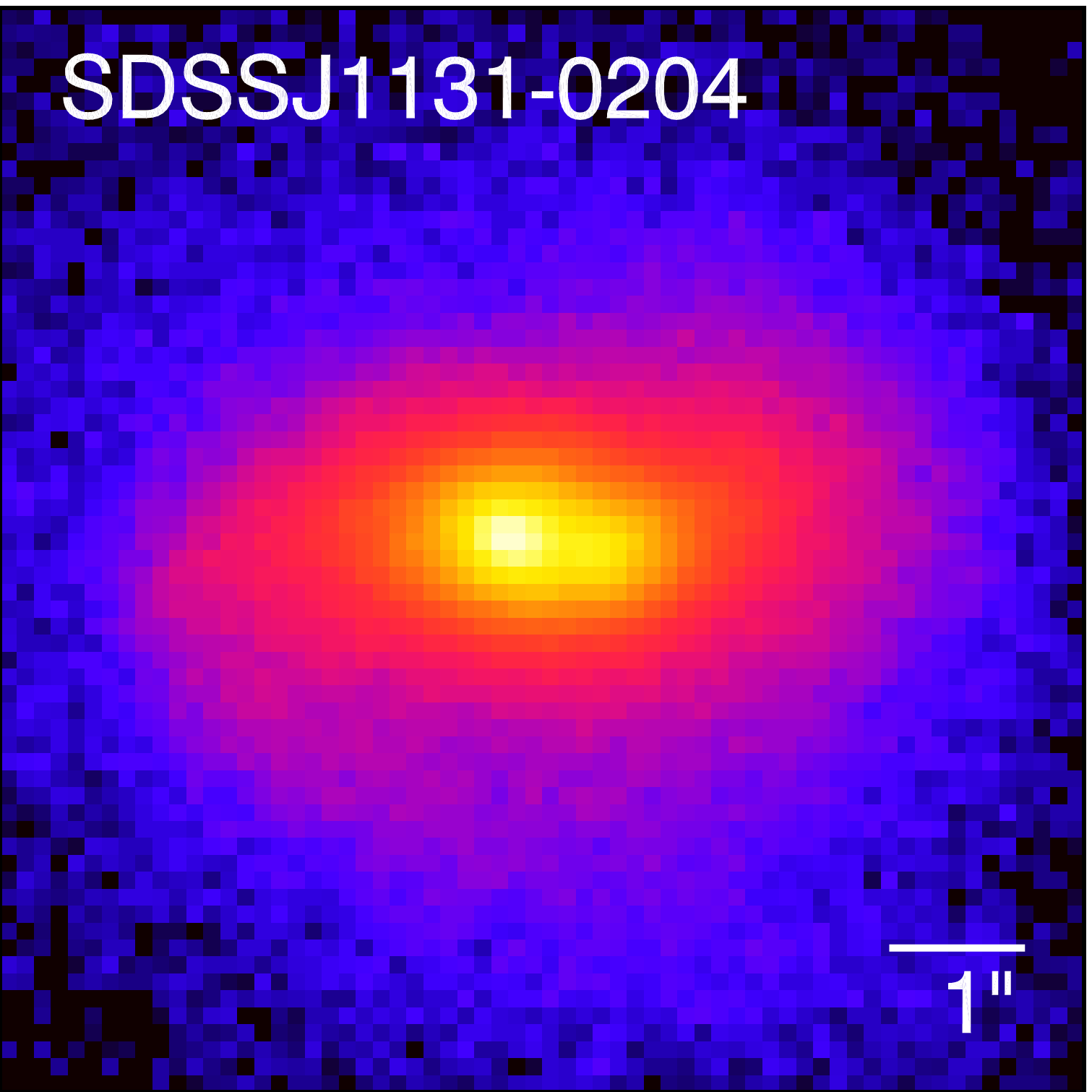}
    \includegraphics[width=40mm]{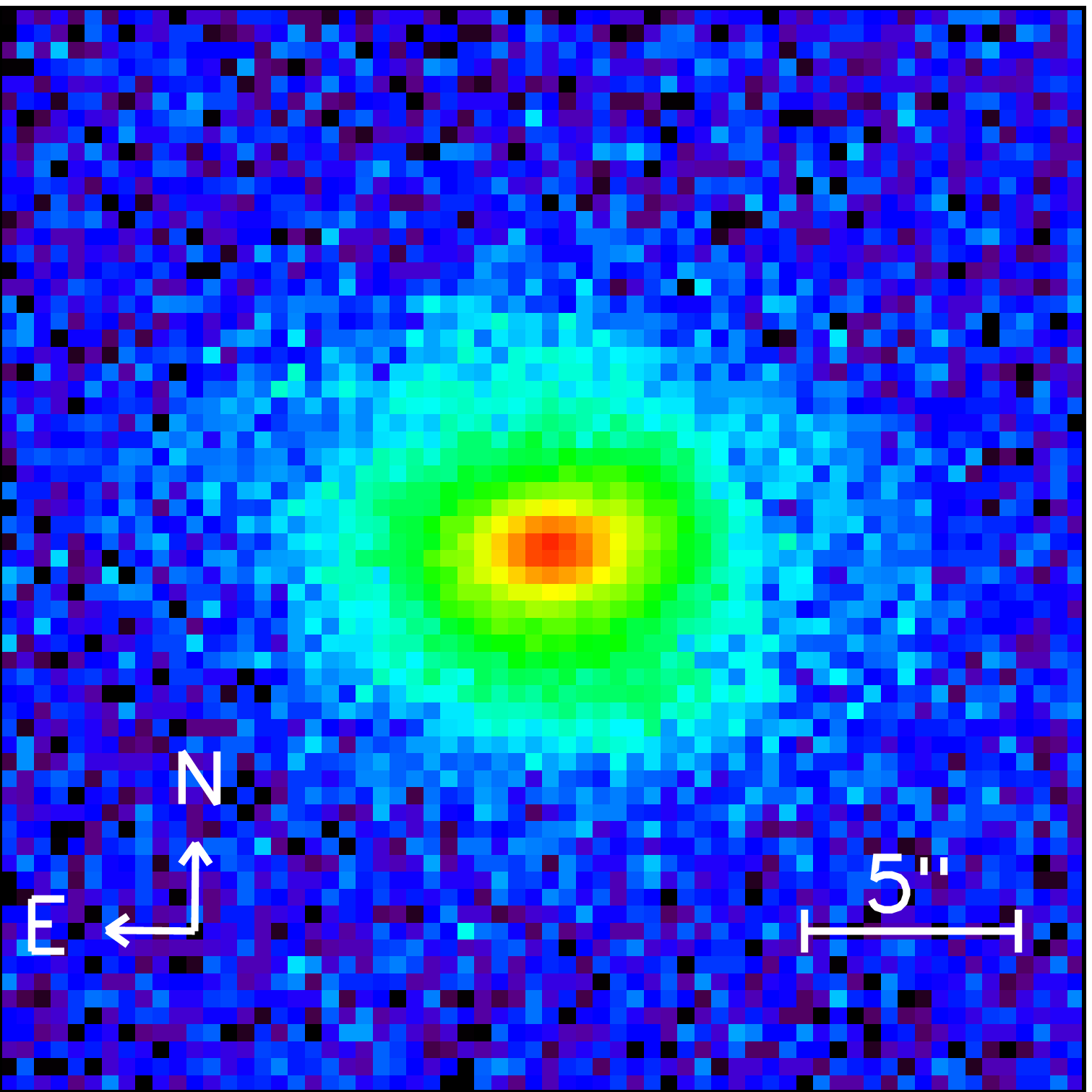}
    \includegraphics[width=40mm]{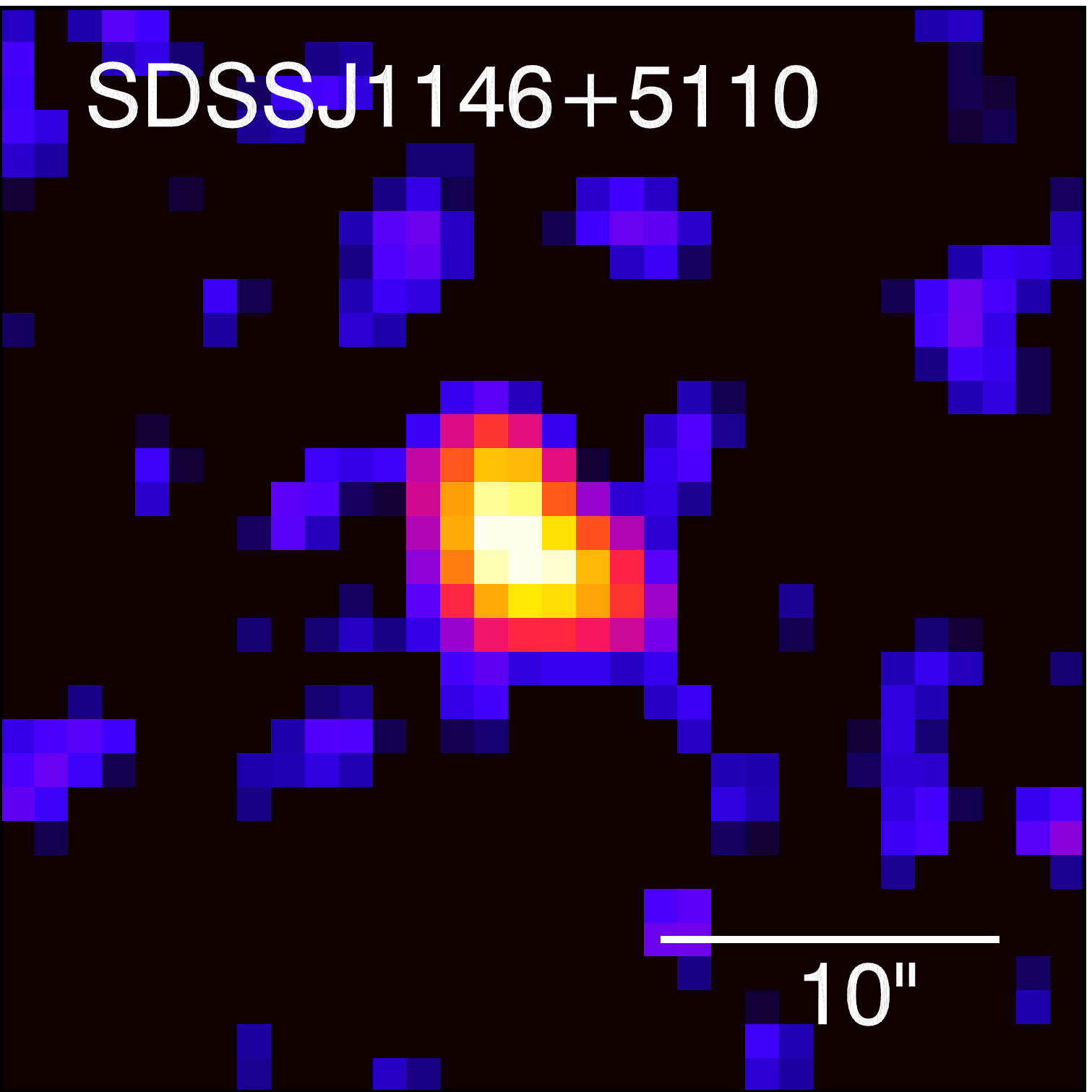}
    \includegraphics[width=40mm]{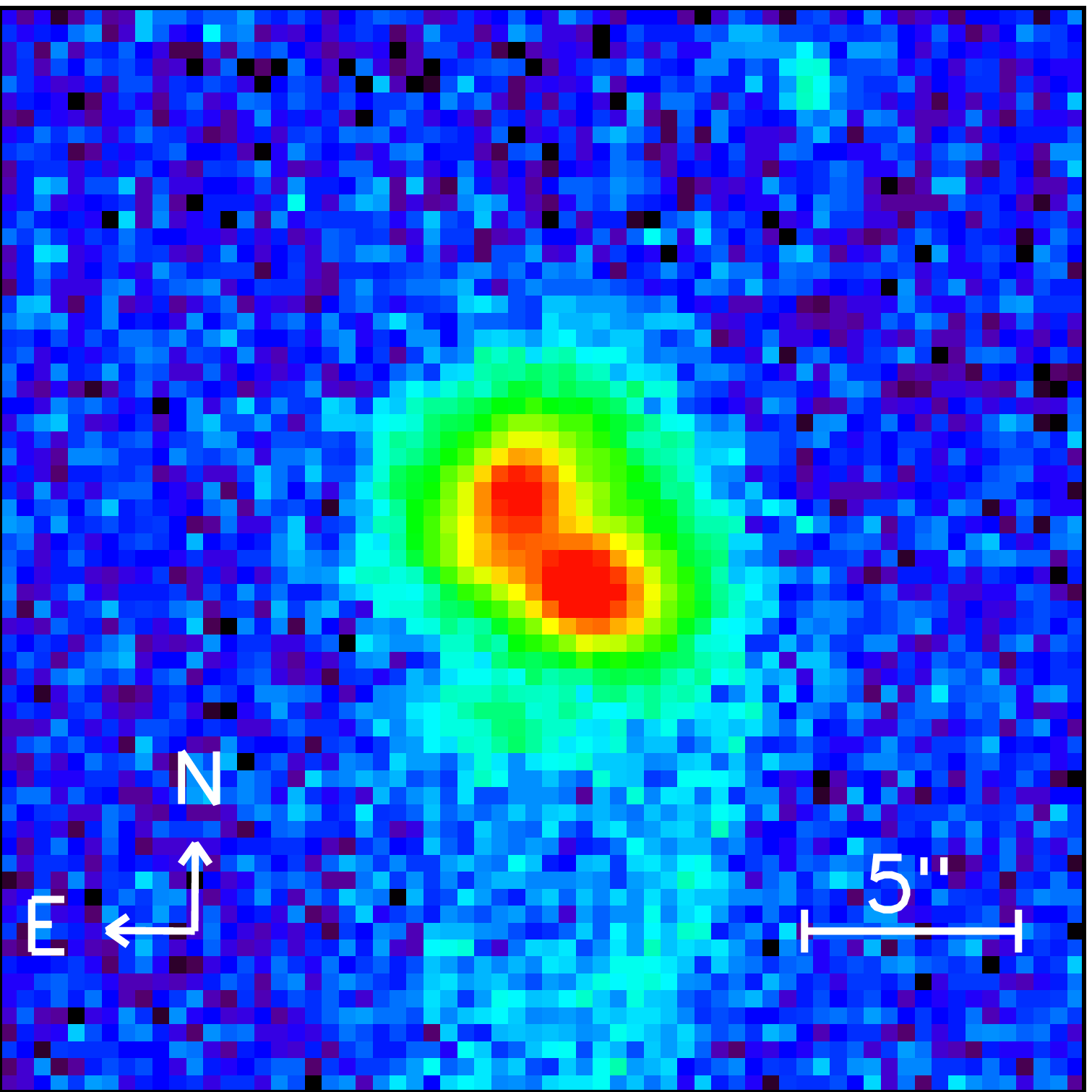}
    \includegraphics[width=40mm]{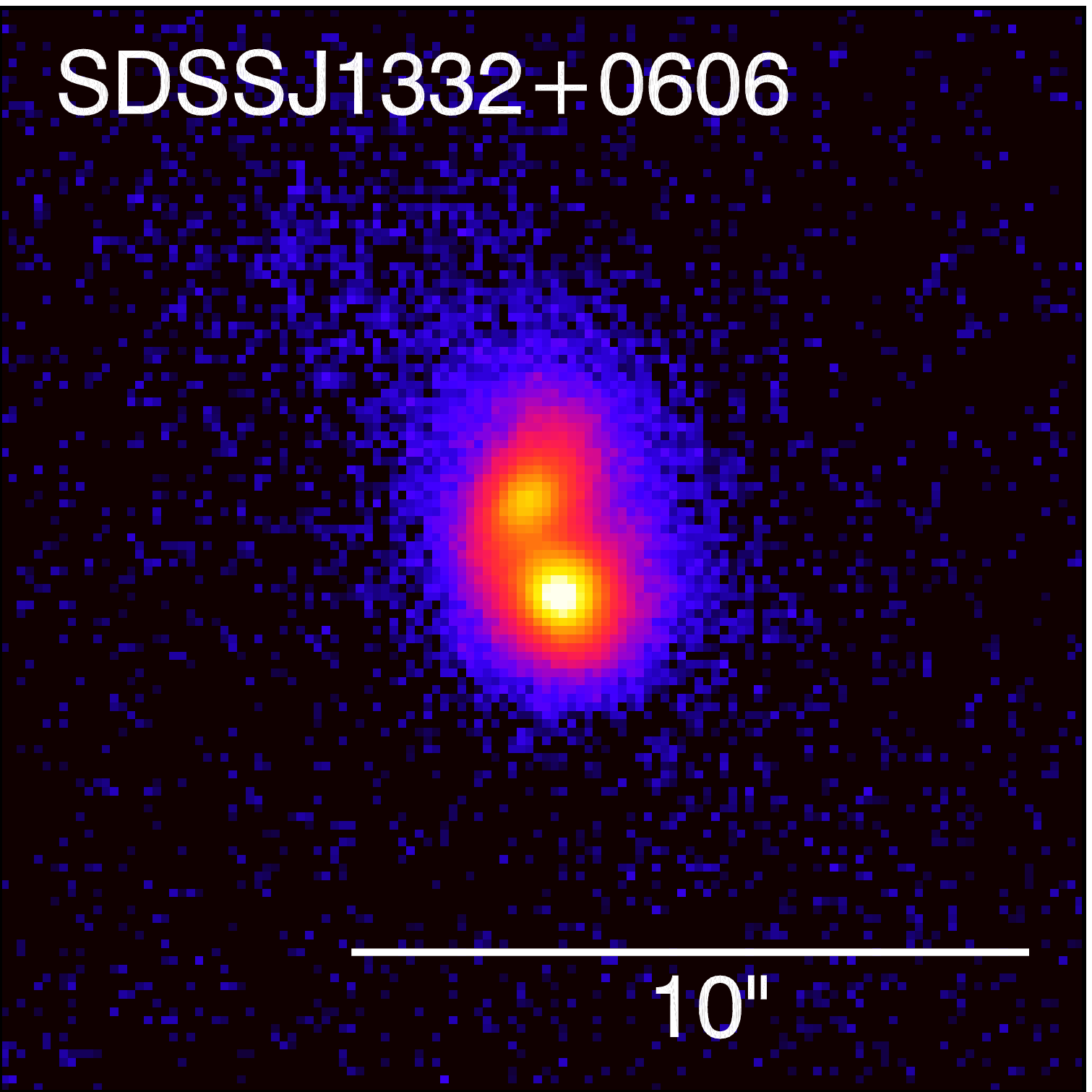}
    \includegraphics[width=40mm]{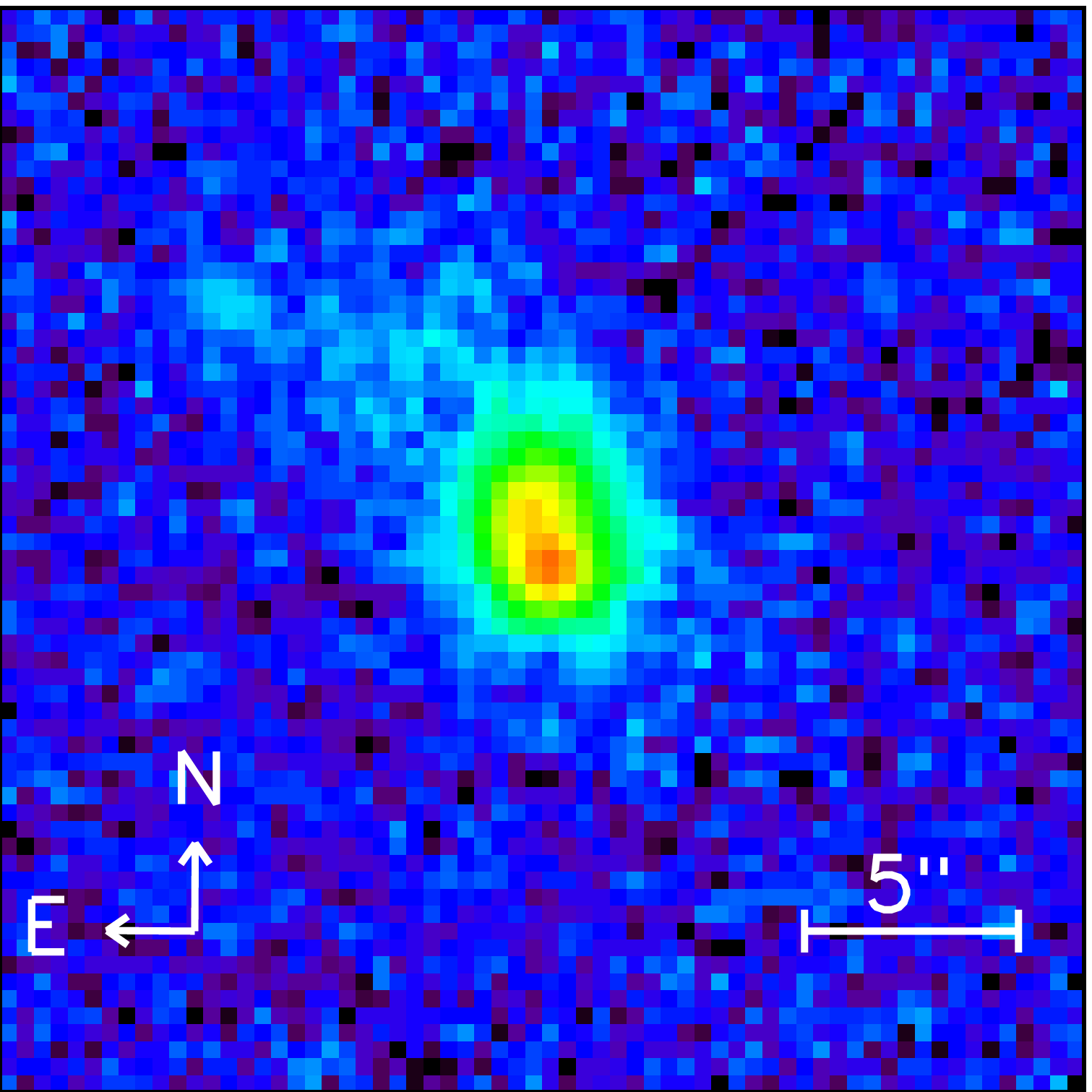}
    \caption{({\it Left}) NIR images for SDSSJ$114642.47+511029.6$
    ($K_s$) from 2MASS and for SDSSJ$110851.04+065901.4$ ($K_s$), SDSSJ$113126.08-020459.2$ ($K_s$), and
    SDSSJ$133226.34+060627.4$ ($J$) from our PANIC observations. North is up and east is to the left.
    Spatial scales are indicated on each panel. Colors are an asinh
    scale to intensity with a stretch to highlight the central nuclei.
    ({\it Right}) SDSS $r$-band images to show tidal
    features and galaxy-scale environment.}
    \label{fig:nirimg}
\end{figure}

We obtained $K_s$ and $J$ images for SDSSJ$113126.08-020459.2$, a
$K_s$ image for SDSSJ$110851.04+065901.4$, and a $J$ image for
SDSSJ$133226.34+060627.4$ on the nights of 2009 December 29
through 2010 January 2 UT using the Persson's Auxiliary Nasmyth
Infrared Camera \citep[PANIC;][]{martini04} on the 6.5 m Magellan
I (Baade) telescope.  PANIC has a $2'\times2'$ field of view (FOV)
and 0\arcsec.126 pixels.  The observing conditions were clear but
not photometric, with seeing ranging between 0\arcsec.5 and
0\arcsec.8 in the optical.  Exposure times were 1080 s and 2160 s
in $K_s$ for SDSSJ$110851.04+065901.4$ and
SDSSJ$113126.08-020459.2$ respectively, and 1080 s in $J$ for both
SDSSJ$113126.08-020459.2$ and SDSSJ$133226.34+060627.4$. We
reduced PANIC data with the Carnegie Supernova Project pipeline
\citep{hamuy06} following standard procedures.

Figure \ref{fig:nirimg} shows the $K_s$ image for
SDSSJ$114642.47+511029.6$ from 2MASS \citep{skrutskie06}, and the
$K_s$ images for SDSSJ$110851.04+065901.4$ and
SDSSJ$113126.08-020459.2$ and the $J$ image for
SDSSJ$133226.34+060627.4$. Also presented are the SDSS $r$-band
images to illustrate the galaxy-scale environment and tidal
features. The NIR images show two stellar bulges in each of these
systems. In addition, all four systems show tidal features
suggestive of mergers.  For SDSSJ$114642.47+511029.6$, this is
clearly seen in its SDSS image; in SDSSJ$110851.04+065901.4$,
SDSSJ$113126.08-020459.2$, and SDSSJ$133226.34+060627.4$, the SDSS
images show tentative tidal features which are confirmed in deep
PANIC images. The PANIC image of SDSSJ$110851.04+065901.4$ also
reveals a small companion ($\sim$3\arcsec west to the nuclei)
which is barely resolved in its SDSS image. The two stellar nuclei
in SDSSJ$110851.04+065901.4$ appear marginally resolved in the
center; their SDSS counterparts exhibit different colors which
could indicate different stellar populations or dust effects.  We
performed aperture photometry for the two stellar nuclei to infer
their luminosity ratio. Table \ref{table:result} lists the
estimated luminosity ratios in $K_s$ or in $J$. Assuming similar
mass-to-light ratios in the two stellar nuclei, they all appear to
be major mergers with comparable stellar masses.   Also listed in
Table \ref{table:result} are the projected separation $\Delta
\theta_{i}$, the stellar mass of the whole galaxy estimated based
on SDSS photometry \citep{kauffmann03}, and the 2MASS $K_s$
magnitudes and $K_s - J$ colors.

\subsection{Optical Slit Spectroscopy}\label{subsec:ldss3}

\begin{figure}
  \centering
    \includegraphics[width=57mm]{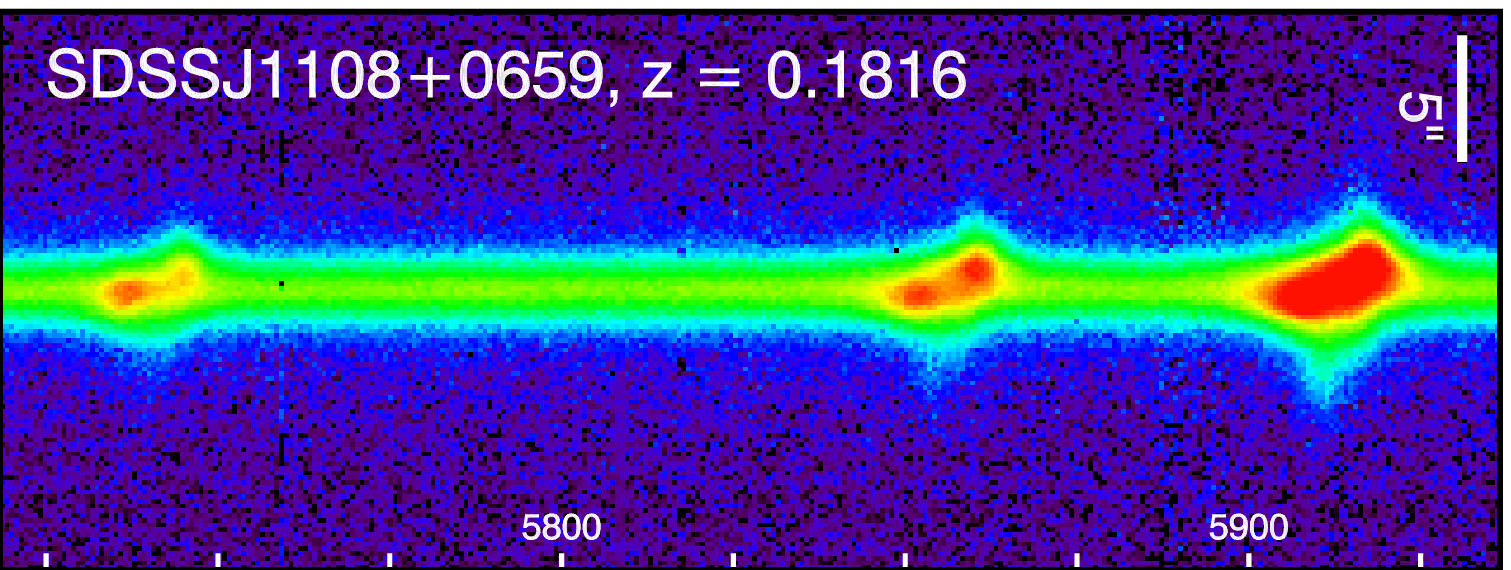}
    \includegraphics[width=22.32mm]{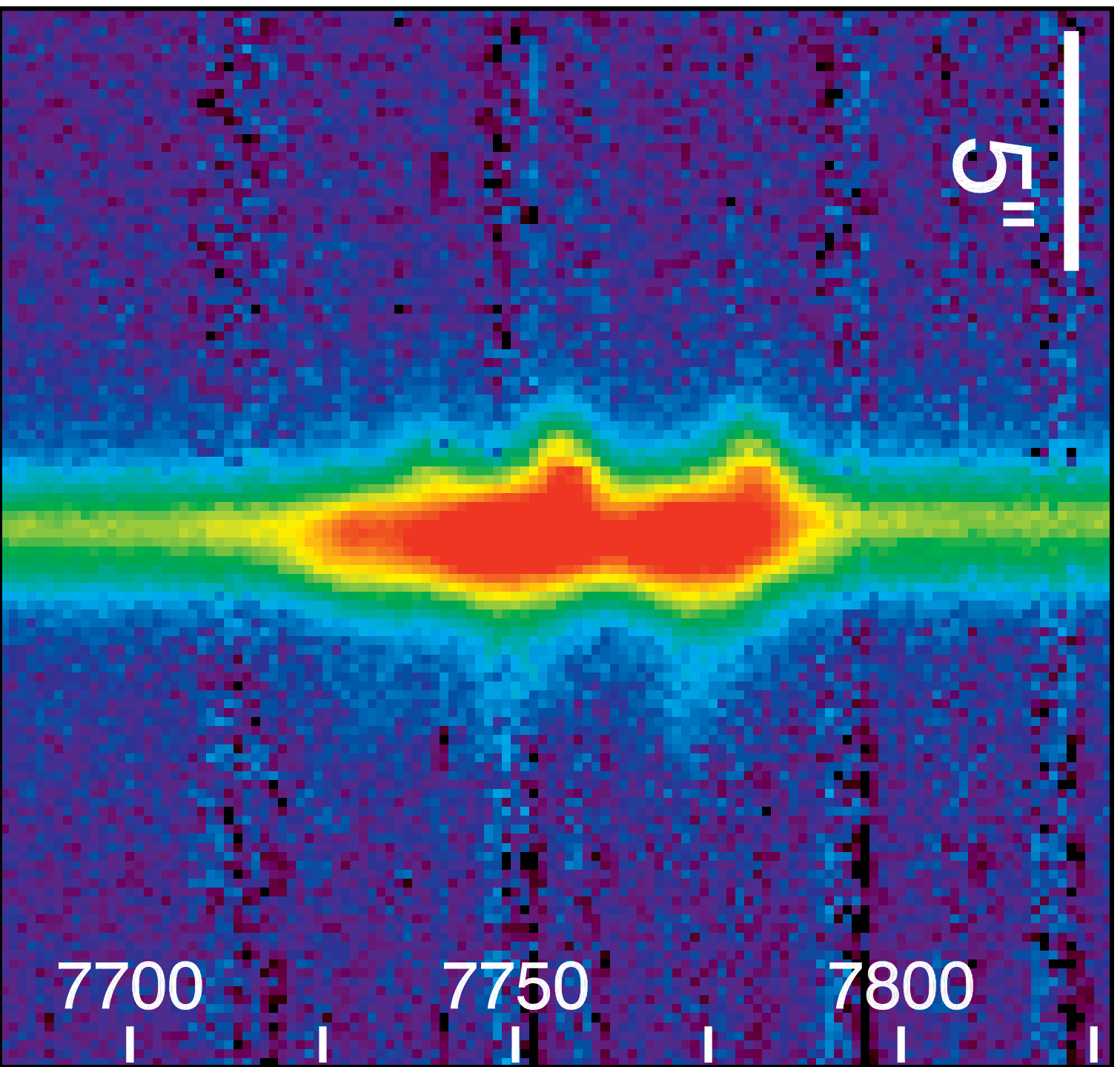}
    \includegraphics[width=57mm]{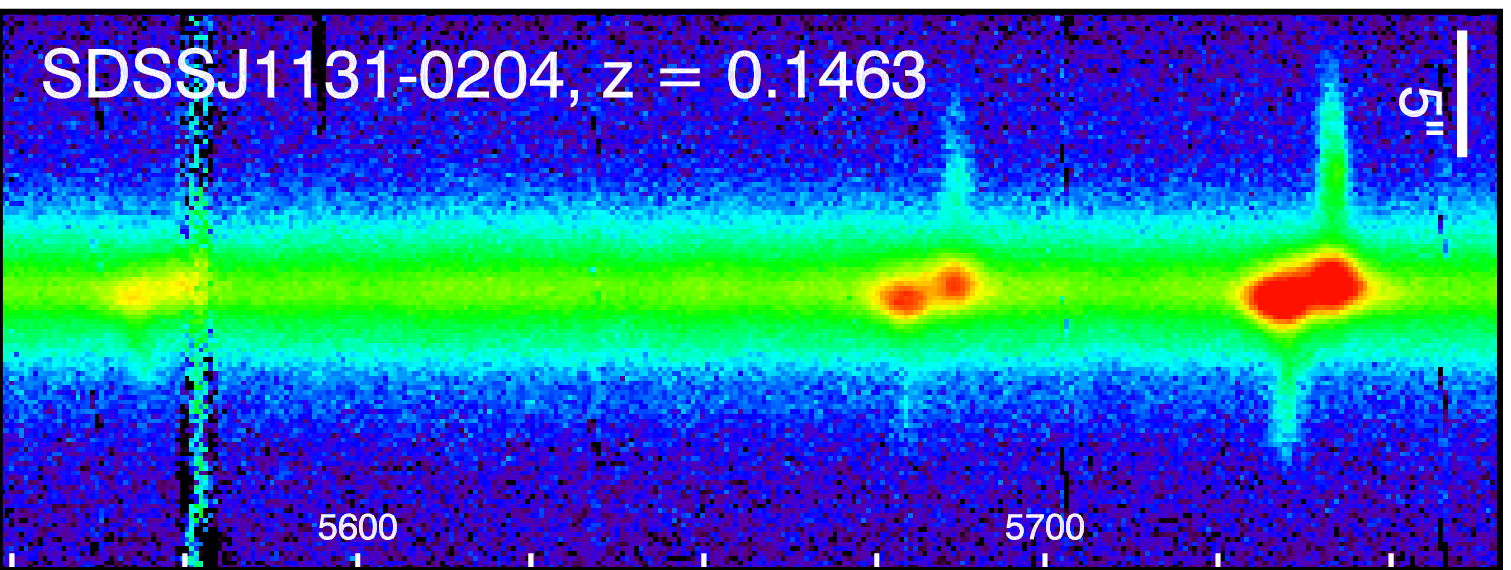}
    \includegraphics[width=22.32mm]{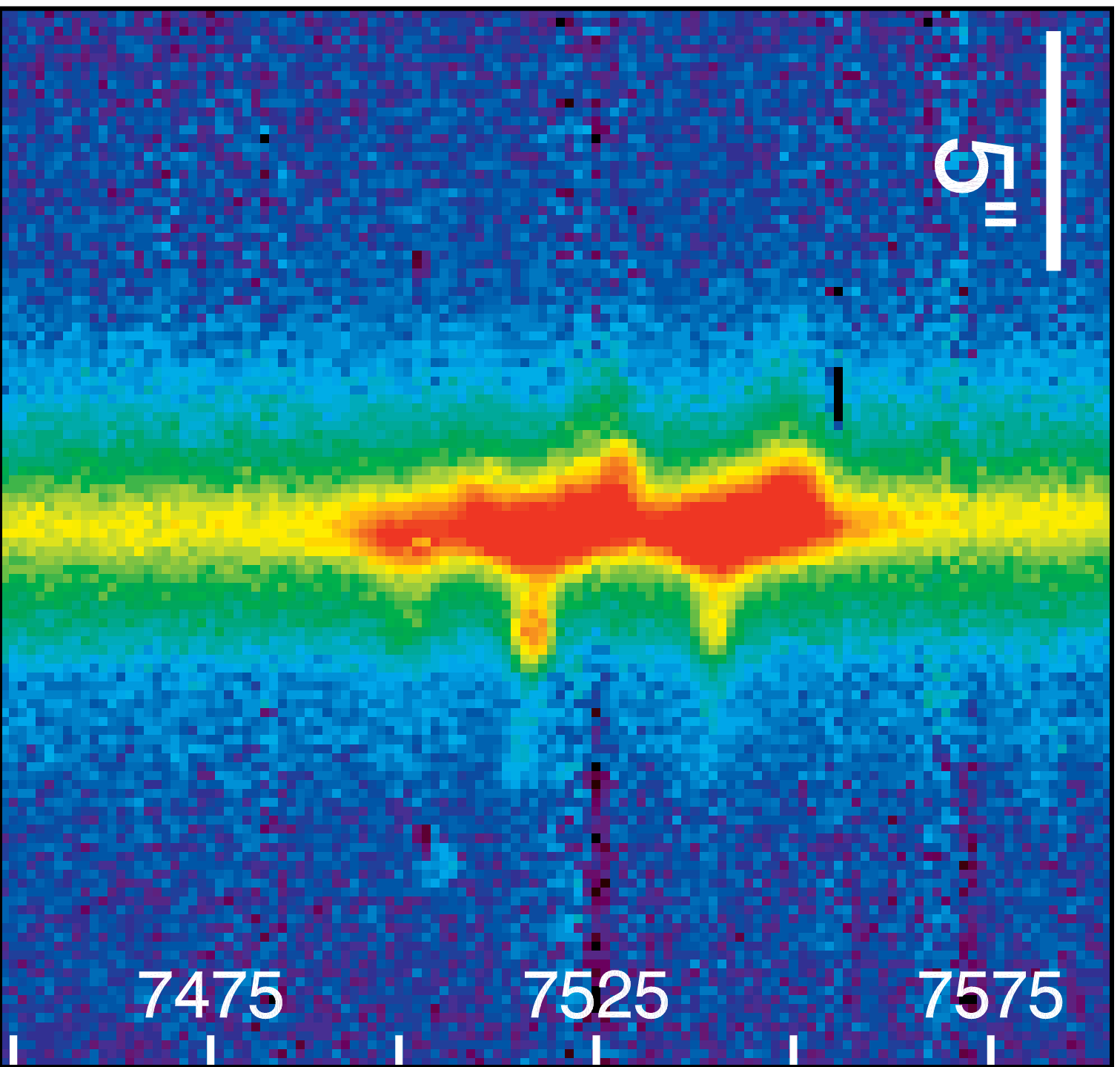}
    \includegraphics[width=57mm]{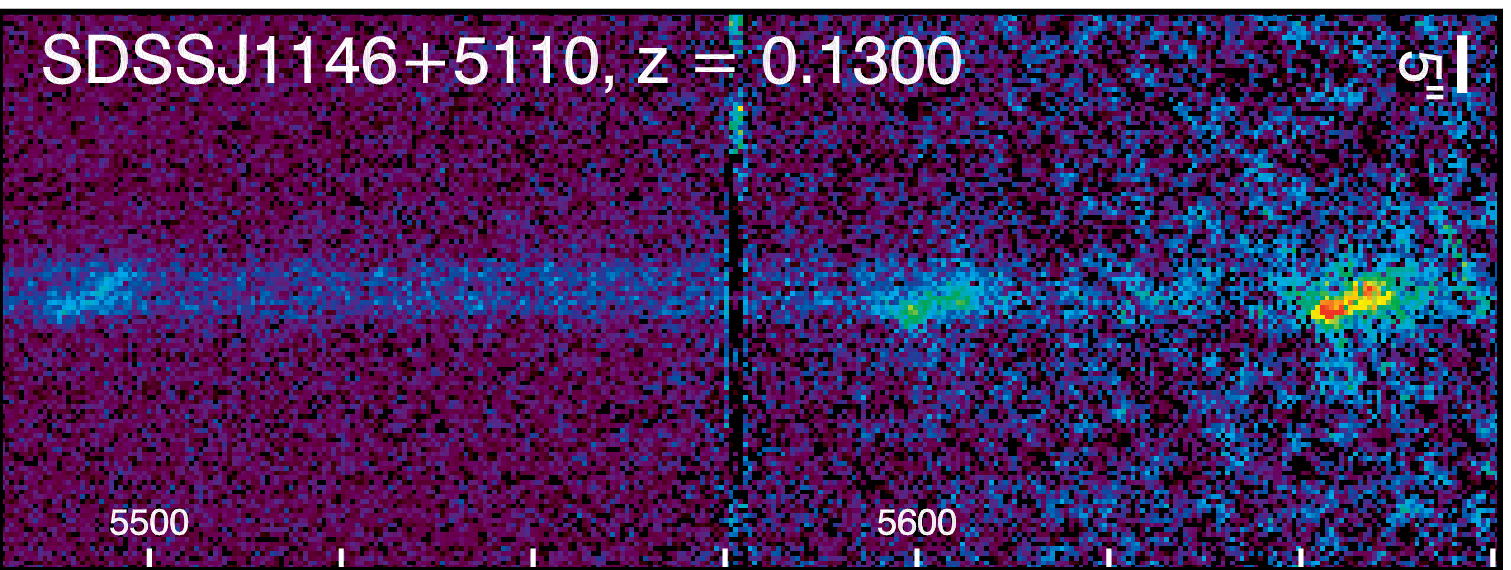}
    \includegraphics[width=22.32mm]{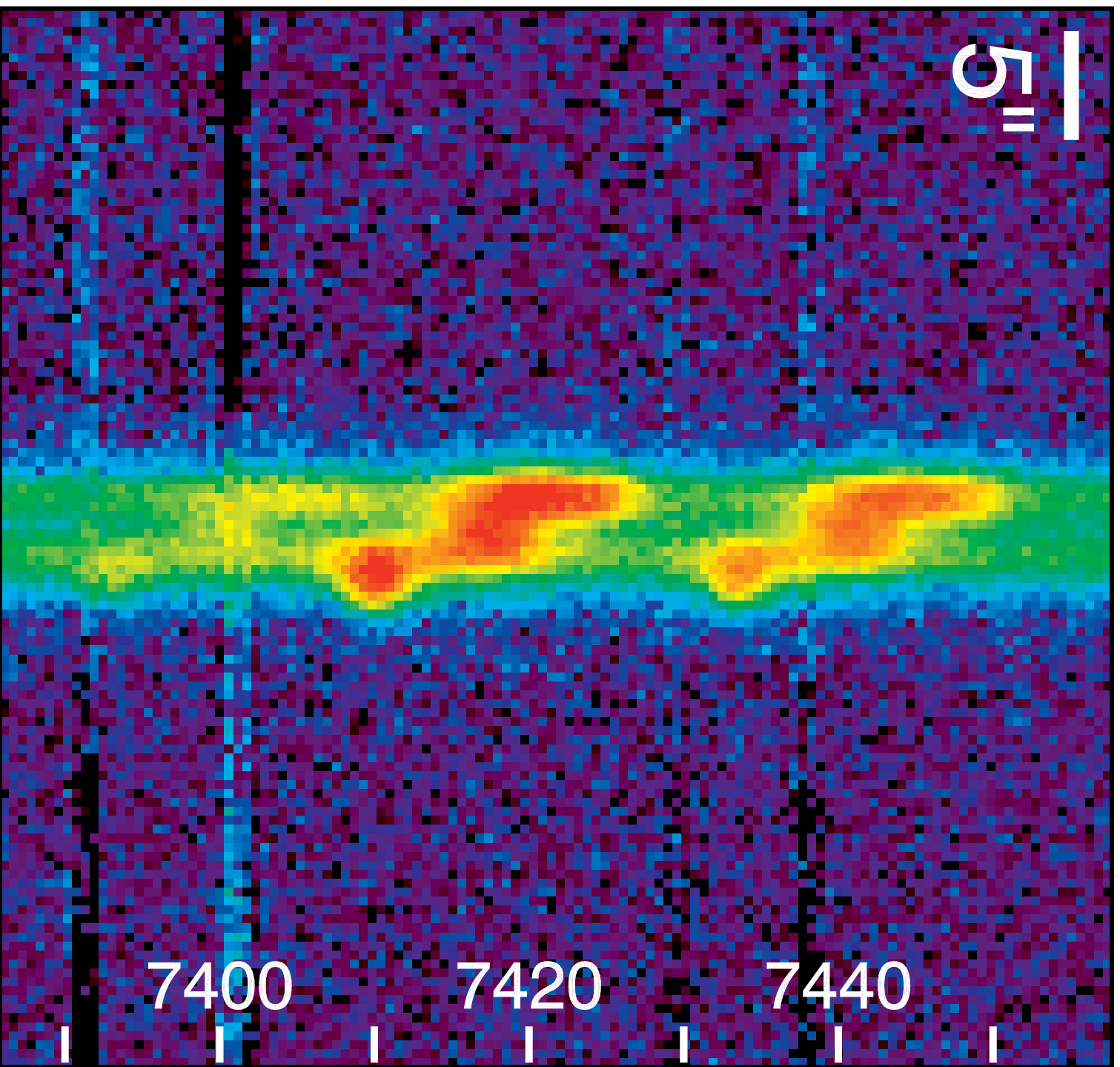}
    \includegraphics[width=57mm]{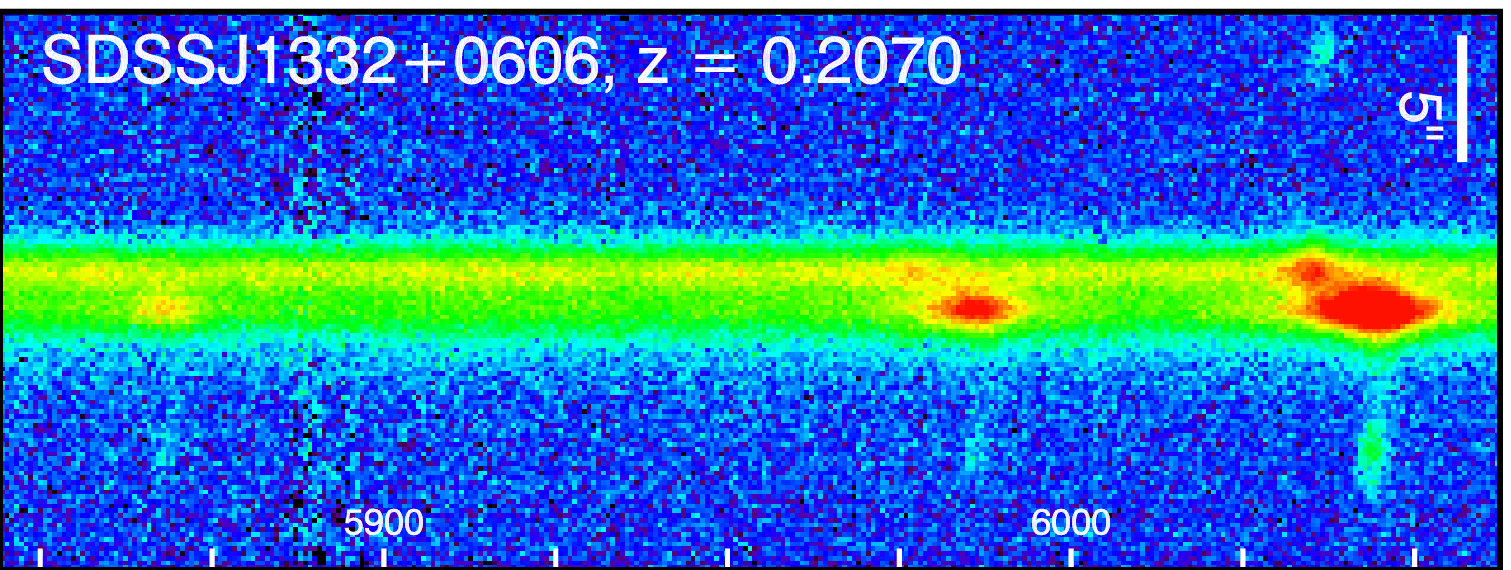}
    \includegraphics[width=22.32mm]{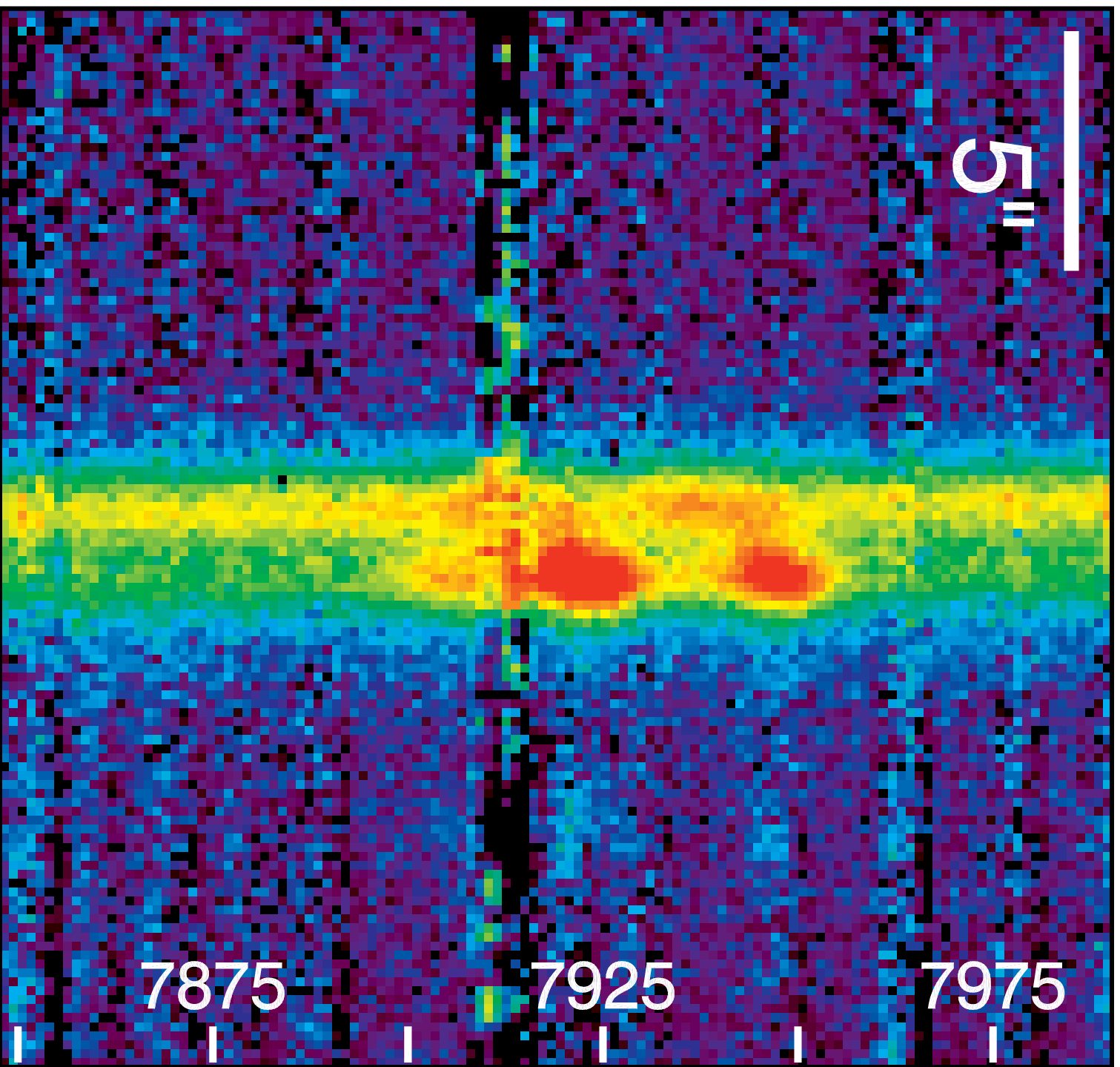}
    \includegraphics[width=57mm]{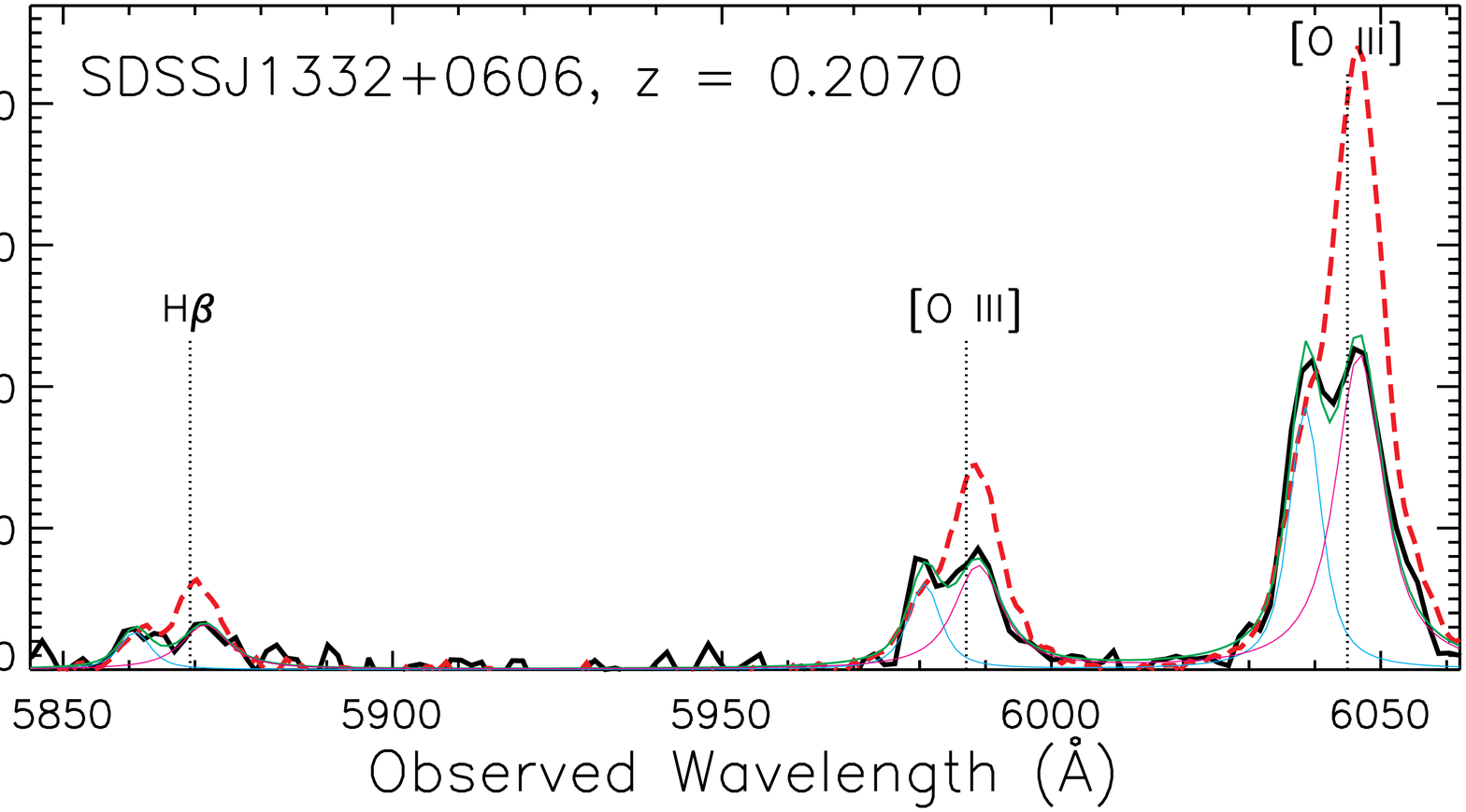}
    \includegraphics[width=22.32mm]{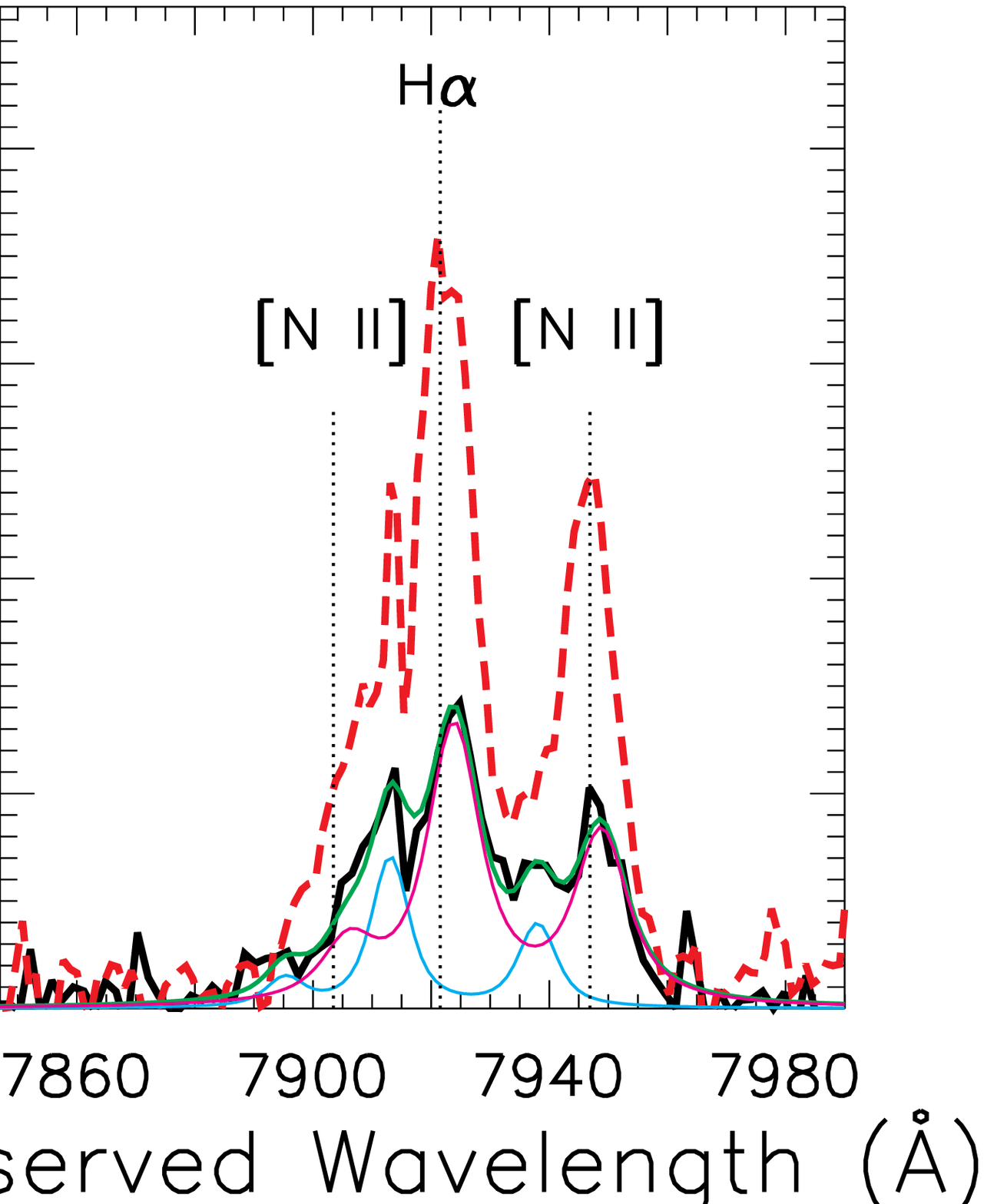}
    \caption{(Top four rows) LDSS3 and DIS 2D spectra over the
    \hbeta-\OIII\ (left column) and \halpha-\NII\
    (right column) regions. Colors indicate flux densities
    with asinh scales and a stretch to illustrate
    both nuclei and extended emission. In both SDSSJ$114642.47+511029.6$ and
    SDSSJ$133226.34+060627.4$, the blueshifted (redshifted) component corresponds
    to the SW (NE) stellar bulge in its image
    (Figure \ref{fig:nirimg}). In SDSSJ$113126.08-020459.2$, the
    blueshifted (redshifted) component corresponds to the E
    (W) bulge in its image, while in SDSSJ$110851.04+065901.4$, the blueshifted
    (redshifted) component corresponds to the NW (SE) bulge
    in its image.
    (Bottom row) Stellar-continuum subtracted spectra for SDSSJ$133226.34+060627.4$
    from LDSS3 (red, dashed curves) and from SDSS
    (data in black, best-fit model in green,
    and the two velocity components in cyan and magenta). The
    vertical dotted lines are drawn at the systemic redshift
    of each object as measured from stellar absorption features.
    In SDSSJ$133226.34+060627.4$ the SDSS fiber covered only part of the
    redshifted component.}
    \label{fig:2dspec}
\end{figure}

Because of the lack of spatial information in the SDSS spectra, it
is unclear whether the two velocity components seen in \OIII\ are
spatially associated with the two stellar bulges.  To resolve this
question we obtained slit spectra for SDSSJ$110851.04+065901.4$,
SDSSJ$113126.08-020459.2$, and SDSSJ$133226.34+060627.4$ on the
nights of 2010 January 12 through 14 UT using the Low-Dispersion
Survey Spectrograph (LDSS3) on the 6.5 m Magellan II (Clay)
telescope. The observing conditions were clear but not
photometric, with seeing ranging between 0\arcsec.7 and
1\arcsec.1. LDSS3 has a 8$'$.3 diameter FOV and 0\arcsec.188
pixels. We employed a 1$''\times$4$'$ long-slit with the VPH-Blue
grism to cover \hbeta\ and \OIIIc , and the VPH-Red grism (with
the OG590 filter) to cover \halpha\ and \NII . We oriented the
slit to go through the two stellar bulges seen in the NIR images
(with PA = 141$^{\circ}$, 80$^{\circ}$, and 17$^{\circ}$ for
SDSSJ$110851.04+065901.4$, SDSSJ$113126.08-020459.2$, and
SDSSJ$133226.34+060627.4$ respectively).  The spectral resolution
was 3.1 (6.3) \angstrom\ FWHM in the blue (red). Total exposure
times were 4800s, 6300s, and 4500s for SDSSJ$110851.04+065901.4$,
SDSSJ$113126.08-020459.2$, and SDSSJ$133226.34+060627.4$
respectively in the blue, and 1800s for all three objects in the
red.  We took wavelength calibration spectra and flat fields after
observing each object.  We obtained spectra for
SDSSJ$114642.47+511029.6$ on the night of 2010 February 15 UT
using the Dual Imaging Spectrograph (DIS) on the Apache Point
Observatory 3.5 m telescope.  The sky was photometric, albeit with
poor seeing ranging between 1\arcsec.7 and 2\arcsec.7. DIS has a
4$'\times$6$'$ FOV and 0\arcsec.414 pixels.  We adopted a
1\arcsec.5$\times$6$'$ slit and the B1200+R1200 gratings centered
at 5500 and 7450 \angstrom . The slit was oriented with PA $=
59^{\circ}$ to go through both bulges shown in the SDSS image. The
spectral resolution was 1.8 (1.3) \angstrom\ FWHM in the blue
(red). The total exposure time was 4800s. We reduced the DIS and
LDSS3 data following standard IRAF\footnote{IRAF is distributed by
the National Optical Astronomy Observatory, which is operated by
the Association of Universities for Research in Astronomy (AURA)
under cooperative agreement with the National Science Foundation.}
procedures \citep{tody86} and with the COSMOS reduction
pipeline\footnote{http://www.ociw.edu/Code/cosmos}. We observed
white dwarfs EG21 and G191B2B for spectrophotometric calibration.
We applied a telluric correction from our standard stars after
extracting 1D spectra.

The spatial correspondence between the two stellar continuum peaks
and two \OIII\ velocity peaks in all cases strongly indicates we
are seeing two active galaxies (Figure \ref{fig:2dspec}).  We see
unambiguous spatial separations between the red and blue velocity
components in all targets and in each strong emission line. We
measure this spatial offset, $\Delta \theta_{e}$, from a spatial
profile summed over twice the FWHM of each line (although we
combine \halpha\ and \NII\ due to blending).  We also checked
these results by directly examining the 2D flux-density profile of
each velocity component.  The spatial offsets derived from each
strong line [\hbeta , \OIIIa , \OIIIb\ ($\Delta \theta_{{\rm [O
\,\,{\scriptscriptstyle III}]}}$ in Table \ref{table:result}), and
the \halpha-\NII\ region ($\Delta \theta_{{\rm H}\alpha}$ in Table
\ref{table:result})] agree within $\sim$0\arcsec.3.  Note in the
case of SDSSJ$114642.47+511029.6$, the \OIIIc\ lines fall in the
dichroic edge and are poorly detected in the redshifted
component\footnote{In SDSSJ$114642.47+511029.6$, the blueshifted
component itself shows double peaks which also appear spatially
offset from one another; our measurements for the blueshifted
component represent the luminosity-weighted average of both peaks.
The redshifted component overlaps with the redshifted peak of the
blueshifted component in velocity space, and we extracted 1D
spectra for the two components separately to disentangle them
properly.}, but the spatial separation is clearly measured in the
\halpha\ and \NII\ lines. The separation between the two velocity
components of the emission lines is consistent with that of the
double stellar bulges ($\Delta \theta_{i}$ in Table
\ref{table:result}). More directly, the emission lines are
coincident with the stellar continua seen in the spectra.  This
spatial coincidence suggests that the double-peaked emission line
profiles in the SDSS spectra of these objects are due to the
orbital motion of the gas associated with the double stellar
bulges.  This is further supported by the tidal features
indicative of mergers seen in all four systems (\S
\ref{subsec:panic}, Figure \ref{fig:nirimg}).

\subsection{The Nature of the Ionizing Sources}\label{subsec:result}

\begin{figure}
  \centering
    \includegraphics[width=80mm]{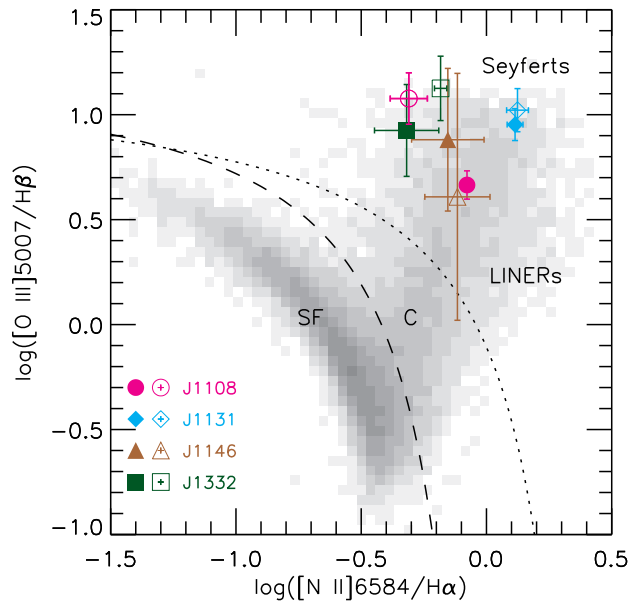}
    \caption{Diagnostic line ratios measured from
    our LDSS3 and DIS spectra.  In each system the two velocity
    components are plotted using the same color,
    with the filled (open) symbol denoting the blueshifted (redshifted) component.
    Gray scales indicate number densities of 31,179
    emission-line galaxies from the SDSS DR4 \citep{kauffmann03}.
    Also shown is the theoretical starburst limit (dotted curve)
    from \citet{kewley01} along with the empirical separation (dashed curve)
    between H {\tiny II} regions and AGNs \citep{kauffmann03}.
    Pure star-forming (SF) galaxies
    lie below the dashed curve, AGN-dominated
    objects lie above the dotted curve, and H {\tiny II}-AGN
    composites (C) lie in between.}
    \label{fig:bpt}
\end{figure}

\begin{deluxetable*}{cccccccccccccccc}
\tabletypesize{\scriptsize} 
\tablecolumns{15}
\tablewidth{0pc} 
\tablecaption{Host-Galaxy and Emission-Line Properties. 
\label{table:result}
} 
\tablehead{ 
\colhead{} &
\colhead{} & 
\colhead{} & 
\colhead{} & 
\colhead{} & 
\colhead{} &
\colhead{$\Delta \theta_{i}$} & 
\colhead{$\Delta S_{i}$} &
\colhead{} & 
\colhead{} & 
\colhead{} &
\colhead{$\Delta\theta_{{\rm [O\,\,{\scriptscriptstyle III}]}}$} &
\colhead{$\Delta \theta_{{\rm H}\alpha}$} & \colhead{} &
\colhead{} \\
\colhead{SDSS Designation} & 
\colhead{Redshift} &
\colhead{$M_{\ast}$} & 
\colhead{$K_s$} & 
\colhead{$g - r$} &
\colhead{$J - K_s$} & 
\colhead{($''$)} & 
\colhead{(kpc)} &
\colhead{$L_1/L_2$} & 
\colhead{$L_{{\rm [O\,\,{\scriptscriptstyle III}]}_1}$} & 
\colhead{$L_{{\rm [O\,\,{\scriptscriptstyle III}]}_2}$} & 
\colhead{($''$)} & \colhead{($''$)} &
\colhead{$n_{e_1}$} &
\colhead{$n_{e_2}$} \\
\colhead{(1)} & 
\colhead{(2)} & 
\colhead{(3)} & 
\colhead{(4)} &
\colhead{(5)} & 
\colhead{(6)} & 
\colhead{(7)} & 
\colhead{(8)} &
\colhead{(9)} & 
\colhead{(10)} & 
\colhead{(11)} & 
\colhead{(12)} &
\colhead{(13)} & 
\colhead{(14)} & 
\colhead{(15)} 
} 
\startdata
SDSSJ$110851.04+065901.4$\dotfill & 0.1816 & 11.0 & 14.29 & 0.8 & 1.3 & 0.5 & 1.5 & 1.0 & 8.58 & 7.94 & 0.9 & 0.6 & 2.4$_{-0.2}^{+0.1}$ & 2.4$_{-1.3}^{+0.7}$  \\
SDSSJ$113126.08-020459.2$\dotfill & 0.1463 & 11.3 & 14.51 & 1.0 & 1.3 & 0.6 & 1.5 & 1.8 & 7.81 & 7.72 & 0.6 & 0.6 & 2.7$_{-0.4}^{+0.2}$ & -0.3$_{-0.7}^{+1.7}$ \\
SDSSJ$114642.47+511029.6$\dotfill & 0.1300 & 10.9 & 14.32 & 1.0 & 1.4 & 2.7 & 6.3 & 0.9 & 8.35 & 7.79 & 2.5 & 2.8 & 3.1$_{-0.4}^{+0.5}$ & 2.9$_{-0.4}^{+1.0}$  \\
SDSSJ$133226.34+060627.4$\dotfill & 0.2070 & 11.1 & 14.72 & 1.2 & 1.6 & 1.5 & 5.1 & 1.6 & 7.48 & 8.25 & 1.6 & 1.6 & 2.8$_{-0.8}^{+0.4}$ & 2.4$_{-0.5}^{+0.3}$  
\enddata
\tablecomments{The subscripts ``1'' and ``2'' denote blueshifted
and redshifted components.  Col. 3: total stellar masses in the
form of log($M_{\ast}/M_{\odot}$) for the whole galaxy based on
SDSS photometry from \citet{kauffmann03}; Col. 4: from 2MASS; Col.
5: color from SDSS photometry; Col. 6: color from 2MASS; Cols. 7
\& 8: projected separation based on NIR imaging; Col. 9: $J$ (for
SDSSJ$133226.34+060627.4$) or $Ks$-band (for the others)
luminosity ratio of the two stellar bulges estimated with aperture
photometry; Cols. 10 \& 11: blueshifted and redshifted \OIIIb\
component luminosities in the form of log($L/L_{\odot}$) ; Col.
12: projected spatial offset measured from \OIIIb ; Col. 13:
projected spatial offset measured from \halpha\ and \NII ; Cols.
14 \& 15: electron density in the form of log($n_e/{\rm cm}^{-3}$)
of the blueshifted and redshifted components.}
\end{deluxetable*}

The nature of the ionizing source in each object is constrained
using the BPT diagram \citep{bpt,veilleux87} featuring the
emission line ratios \OIIIb/\hbeta\ and \NIIb/\halpha\ (Figure
\ref{fig:bpt}).  We fit multi-component models to the emission
lines over the stellar-continuum subtracted LDSS3 and DIS spectra
\citep[refer to][for details on constructing continuum models and
emission line fitting]{liu10} and measured line ratios.  For
SDSSJ$110851.04+065901.4$, SDSSJ$113126.08-020459.2$, and
SDSSJ$133226.34+060627.4$, we extracted 1D spectra summing over
both velocity components as they are partially blended spatially,
whereas for SDSSJ$114642.47+511029.6$ we extracted 1D spectra
separately for each of the two velocity components. We compared
measurements based on the SDSS and the LDSS3 and DIS spectra and
found general agreement in the line ratios; the difference in line
fluxes is likely due to the difference in the fiber and slit
coverage. Figure \ref{fig:2dspec} (bottom panels) shows an example
of our emission line fits and of the difference in SDSS and slit
spectra. In all cases, the diagnostic line ratios indicate
Seyfert-type ionization (except for the redshifted component in
SDSSJ$114642.47+511029.6$, which could also be a LINER or a
composite). Table \ref{table:result} lists \OIIIb\ luminosities
for both velocity components based on our LDSS3/DIS spectra.

While the diagnostic line ratios indicate the presence of hard
ionizing photons, it is ambiguous whether there are two ionizing
sources.  If only one nucleus were to ionize both gas components,
the ionization parameter in the off-nuclear region would be much
lower than that in the nuclear region with similar electron
densities, because the number density of ionizing photons decays
as $r^{-2}$.  We measured electron densities $n_e$ using the
diagnostic line ratio \SIIa/\SIIb .  For
SDSSJ$110851.04+065901.4$, SDSSJ$114642.47+511029.6$, and
SDSSJ$133226.34+060627.4$, we found comparable $n_e$ values in the
two velocity components (Table \ref{table:result}).  The
comparable electron densities and high ionization states (Figure
\ref{fig:bpt}) in both narrow line components suggest that there
are separate hard ionizing sources associated with each line
component.  For SDSSJ$113126.08-020459.2$, the inferred $n_e$ of
the redshifted component is $\sim$1000 times smaller than the
blueshifted component, but the uncertainties of the measurement
are so large that they may differ by no more than a factor of
$\sim$ 10. Considering typical $n_e$ values of narrow-line region
gas in stellar bulges, in SDSSJ$113126.08-020459.2$ the comparable
and high \OIIIb/\hbeta\ ratios may still argue for separate
ionizing sources, but it is possible that the nucleus associated
with the blueshifted component is ionizing both velocity
components. Chandra observations would clarify the situation
considerably.  To summarize, our imaging and spectroscopy data
strongly support the binary AGN scenario for at least three of the
four objects.

\section{Discussion}\label{sec:discuss}

As listed in Table \ref{table:result}, the four new kpc-scale
binary AGNs that we have presented here have optical colors
typical of type 2 AGNs \citep[e.g.,][]{kauffmann03}.  All four
objects show tidal features suggestive of mergers (\S
\ref{subsec:panic}, Figure \ref{fig:nirimg}).  Among them,
SDSSJ$110851.04+065901.4$ was detected (but unresolved) in FIRST
with an integrated flux of $9.84\pm0.13$ mJy \citep{white97}; it
also has a candidate match in the IRAS Faint Source Catalog
\citep{moshir92}, IRAS F11062+0715, with $L_{{\rm IR}}\sim 10^{12}
L_{\odot}$, although the positional match is only marginal
($\sim$9.5\arcsec offset from its SDSS position).
SDSSJ$113126.08-020459.2$, SDSSJ$114642.47+511029.6$, and
SDSSJ$133226.34+060627.4$ were not detected in FIRST or in IRAS.
At least three of the four galaxies appear to be neither
star-bursting luminous IR galaxies (LIRGs) nor passively evolving
ellipticals.  These properties are in contrast to the few existing
samples of known kpc-scale binary AGNs, which tend to be found
either in LIRGs which involve gas-rich mergers of disks (NGC 6240,
\citealt{komossa03}; Mrk 463, \citealt{bianchi08}; Arp 299,
\citealt{ballo04}), or in ellipticals at the centers of galaxy
clusters and power twin jets (3C 75, \citealt{owen85}; PKS
2149-158, \citealt{guidetti08}). Perhaps this reflects our
differing selecting criteria, but we are hopeful that with our
systematic approach we can begin to characterize the evolutionary
state of kpc-scale binary AGNs in more detail.  Further host
galaxy studies on these four binary AGNs including their
structural properties, stellar populations, and dynamics, as well
as the remainder of our NIR imaging subsample will be presented
elsewhere.

Until recently, searches for kpc-scale binary AGNs have been more
or less serendipitous.  We have demonstrated with the observations
presented in this paper that selecting spectroscopic candidates
based on double-peaked narrow lines and follow-up with high
spatial resolution imaging and spectroscopy is a promising
technique to identify genuine binary AGNs in an efficient and
systematic way.  Out of the 43 objects observed so far, we
detected three, and perhaps four binary AGNs ($\sim$ 7--10\%).
These four systems probe the relatively early stages of AGN
pairing, when the two stellar bulges which host the two SMBHs are
still separated enough to show two distinct components in
IR/optical images.  The $\sim10\%$ lower limit translates into a
0.1\% kpc-scale binary AGN fraction out of the parent AGN sample
\citep[167 double-peaked objects selected from 14,756
AGNs][]{liu10}, comparable to the binary quasar fraction with
projected separations of tens of kpc
\citep{hennawi06,myers08,hennawi09}, albeit with different
systematics and potential selection biases. We also have objects
in our sample in which the \OIII\ emission is spatially resolved
but the NIR continuum is not, and in a few such cases the
spatially-resolved spectrum shows clear signs of a velocity
gradient across the slit, suggesting a rotation/outflow origin for
the double-peaked profile.  Thus it is premature to claim a binary
AGN based on spatially-resolved \OIII\ emission alone.  A detailed
analysis on the fraction of such ``false positives'' and their
properties will be presented in a future paper.

We are carrying out NIR imaging of the rest of our double-peaked
narrow-line sample using Magellan and Keck with AO.  Together with
follow-up spectroscopy, we expect to increase the number of known
kpc-scale binary AGNs several fold.  High-quality multi-band data
will allow us to explore the properties of individual binary AGNs
in great detail, and the increased statistics will provide
important constraints on the merger hypothesis, AGN lifetimes, and
the interplay between SMBHs and their hosts.

\acknowledgments

We thank J. Krolik for interesting comments, and an anonymous 
referee for a prompt report. X.L. and M.A.S.
acknowledge the support of NSF grant AST-0707266. Y.S.
acknowledges support from a Clay Postdoctoral Fellowship through
the Smithsonian Astrophysical Observatory.

Funding for the SDSS and SDSS-II has been provided by the Alfred
P. Sloan Foundation, the Participating Institutions, the National
Science Foundation, the U.S. Department of Energy, the National
Aeronautics and Space Administration, the Japanese Monbukagakusho,
the Max Planck Society, and the Higher Education Funding Council
for England. The SDSS Web Site is http://www.sdss.org/.


Facilities: Sloan, Magellan: Baade (PANIC), APO ARC 3.5m (DIS)

\bibliography{binaryrefs}

\begin{thebibliography}{44}
\expandafter\ifx\csname natexlab\endcsname\relax\def\natexlab#1{#1}\fi

\bibitem[{{Abazajian} {et~al.}(2009){Abazajian}, {Adelman-McCarthy},
  {Ag{\"u}eros}, {Allam}, {Allende Prieto}, {An}, {Anderson}, {Anderson},
  {Annis}, {Bahcall}, {Bailer-Jones}, {Barentine}, {Bassett}, {Becker},
  {Beers}, {Bell}, {Belokurov}, {Berlind}, {Berman}, {Bernardi}, {Bickerton},
  {Bizyaev}, {Blakeslee}, {Blanton}, {Bochanski}, {Boroski}, {Brewington},
  {Brinchmann}, {Brinkmann}, {Brunner}, {Budav{\'a}ri}, {Carey}, {Carliles},
  {Carr}, {Castander}, {Cinabro}, {Connolly}, {Csabai}, {Cunha}, {Czarapata},
  {Davenport}, {de Haas}, {Dilday}, {Doi}, {Eisenstein}, {Evans}, {Evans},
  {Fan}, {Friedman}, {Frieman}, {Fukugita}, {G{\"a}nsicke}, {Gates},
  {Gillespie}, {Gilmore}, {Gonzalez}, {Gonzalez}, {Grebel}, {Gunn},
  {Gy{\"o}ry}, {Hall}, {Harding}, {Harris}, {Harvanek}, {Hawley}, {Hayes},
  {Heckman}, {Hendry}, {Hennessy}, {Hindsley}, {Hoblitt}, {Hogan}, {Hogg},
  {Holtzman}, {Hyde}, {Ichikawa}, {Ichikawa}, {Im}, {Ivezi{\'c}}, {Jester},
  {Jiang}, {Johnson}, {Jorgensen}, {Juri{\'c}}, {Kent}, {Kessler}, {Kleinman},
  {Knapp}, {Konishi}, {Kron}, {Krzesinski}, {Kuropatkin}, {Lampeitl},
  {Lebedeva}, {Lee}, {Lee}, {Leger}, {L{\'e}pine}, {Li}, {Lima}, {Lin}, {Long},
  {Loomis}, {Loveday}, {Lupton}, {Magnier}, {Malanushenko}, {Malanushenko},
  {Mandelbaum}, {Margon}, {Marriner}, {Mart{\'{\i}}nez-Delgado}, {Matsubara},
  {McGehee}, {McKay}, {Meiksin}, {Morrison}, {Mullally}, {Munn}, {Murphy},
  {Nash}, {Nebot}, {Neilsen}, {Newberg}, {Newman}, {Nichol}, {Nicinski},
  {Nieto-Santisteban}, {Nitta}, {Okamura}, {Oravetz}, {Ostriker}, {Owen},
  {Padmanabhan}, {Pan}, {Park}, {Pauls}, {Peoples}, {Percival}, {Pier}, {Pope},
  {Pourbaix}, {Price}, {Purger}, {Quinn}, {Raddick}, {Fiorentin}, {Richards},
  {Richmond}, {Riess}, {Rix}, {Rockosi}, {Sako}, {Schlegel}, {Schneider},
  {Scholz}, {Schreiber}, {Schwope}, {Seljak}, {Sesar}, {Sheldon}, {Shimasaku},
  {Sibley}, {Simmons}, {Sivarani}, {Smith}, {Smith}, {Smol{\v c}i{\'c}},
  {Snedden}, {Stebbins}, {Steinmetz}, {Stoughton}, {Strauss}, {Subba Rao},
  {Suto}, {Szalay}, {Szapudi}, {Szkody}, {Tanaka}, {Tegmark}, {Teodoro},
  {Thakar}, {Tremonti}, {Tucker}, {Uomoto}, {Vanden Berk}, {Vandenberg},
  {Vidrih}, {Vogeley}, {Voges}, {Vogt}, {Wadadekar}, {Watters}, {Weinberg},
  {West}, {White}, {Wilhite}, {Wonders}, {Yanny}, {Yocum}, {York}, {Zehavi},
  {Zibetti}, \& {Zucker}}]{SDSSDR7}
{Abazajian}, K.~N., {et~al.} 2009, \apjs, 182, 543

\bibitem[{{Axon} {et~al.}(1998){Axon}, {Marconi}, {Capetti}, {Maccetto},
  {Schreier}, \& {Robinson}}]{axon98}
{Axon}, D.~J., {Marconi}, A., {Capetti}, A., {Maccetto}, F.~D., {Schreier}, E.,
  \& {Robinson}, A. 1998, \apjl, 496, L75

\bibitem[{{Baldwin} {et~al.}(1981){Baldwin}, {Phillips}, \& {Terlevich}}]{bpt}
{Baldwin}, J.~A., {Phillips}, M.~M., \& {Terlevich}, R. 1981, \pasp, 93, 5

\bibitem[{{Ballo} {et~al.}(2004){Ballo}, {Braito}, {Della Ceca}, {Maraschi},
  {Tavecchio}, \& {Dadina}}]{ballo04}
{Ballo}, L., {Braito}, V., {Della Ceca}, R., {Maraschi}, L., {Tavecchio}, F.,
  \& {Dadina}, M. 2004, \apj, 600, 634

\bibitem[{{Begelman} {et~al.}(1980){Begelman}, {Blandford}, \&
  {Rees}}]{begelman80}
{Begelman}, M.~C., {Blandford}, R.~D., \& {Rees}, M.~J. 1980, \nat, 287, 307

\bibitem[{{Bianchi} {et~al.}(2008){Bianchi}, {Chiaberge}, {Piconcelli},
  {Guainazzi}, \& {Matt}}]{bianchi08}
{Bianchi}, S., {Chiaberge}, M., {Piconcelli}, E., {Guainazzi}, M., \& {Matt},
  G. 2008, \mnras, 386, 105

\bibitem[{{Comerford} {et~al.}(2009{\natexlab{a}}){Comerford}, {Gerke},
  {Newman}, {Davis}, {Yan}, {Cooper}, {Faber}, {Koo}, {Coil}, {Rosario}, \&
  {Dutton}}]{comerford08}
{Comerford}, J.~M., {et~al.} 2009{\natexlab{a}}, \apj, 698, 956

\bibitem[{{Comerford} {et~al.}(2009{\natexlab{b}}){Comerford}, {Griffith},
  {Gerke}, {Cooper}, {Newman}, {Davis}, \& {Stern}}]{comerford09}
{Comerford}, J.~M., {Griffith}, R.~L., {Gerke}, B.~F., {Cooper}, M.~C.,
  {Newman}, J.~A., {Davis}, M., \& {Stern}, D. 2009{\natexlab{b}}, \apjl, 702,
  L82

\bibitem[{{Crenshaw} {et~al.}(2010){Crenshaw}, {Schmitt}, {Kraemer},
  {Mushotzky}, \& {Dunn}}]{crenshaw09}
{Crenshaw}, D.~M., {Schmitt}, H.~R., {Kraemer}, S.~B., {Mushotzky}, R.~F., \&
  {Dunn}, J.~P. 2010, \apj, 708, 419

\bibitem[{{Davis} {et~al.}(2003){Davis}, {Faber}, {Newman}, {Phillips},
  {Ellis}, {Steidel}, {Conselice}, {Coil}, {Finkbeiner}, {Koo}, {Guhathakurta},
  {Weiner}, {Schiavon}, {Willmer}, {Kaiser}, {Luppino}, {Wirth}, {Connolly},
  {Eisenhardt}, {Cooper}, \& {Gerke}}]{davis03}
{Davis}, M., {et~al.} 2003, in Society of Photo-Optical Instrumentation
  Engineers (SPIE) Conference Series, ed. P.~{Guhathakurta}, Vol. 4834,
  161--172

\bibitem[{{Faber} {et~al.}(1997){Faber}, {Tremaine}, {Ajhar}, {Byun},
  {Dressler}, {Gebhardt}, {Grillmair}, {Kormendy}, {Lauer}, \&
  {Richstone}}]{faber97}
{Faber}, S.~M., {et~al.} 1997, \aj, 114, 1771

\bibitem[{{Gerke} {et~al.}(2007){Gerke}, {Newman}, {Lotz}, {Yan}, {Barmby},
  {Coil}, {Conselice}, {Ivison}, {Lin}, {Koo}, {Nandra}, {Salim}, {Small},
  {Weiner}, {Cooper}, {Davis}, {Faber}, \& {Guhathakurta}}]{gerke07}
{Gerke}, B.~F., {et~al.} 2007, \apjl, 660, L23

\bibitem[{{Green} {et~al.}(2010){Green}, {Myers}, {Barkhouse}, {Mulchaey},
  {Bennert}, {Cox}, \& {Aldcroft}}]{green10}
{Green}, P.~J., {Myers}, A.~D., {Barkhouse}, W.~A., {Mulchaey}, J.~S.,
  {Bennert}, V.~N., {Cox}, T.~J., \& {Aldcroft}, T.~L. 2010, \apj, 710, 1578

\bibitem[{{Guidetti} {et~al.}(2008){Guidetti}, {Murgia}, {Govoni}, {Parma},
  {Gregorini}, {de Ruiter}, {Cameron}, \& {Fanti}}]{guidetti08}
{Guidetti}, D., {Murgia}, M., {Govoni}, F., {Parma}, P., {Gregorini}, L., {de
  Ruiter}, H.~R., {Cameron}, R.~A., \& {Fanti}, R. 2008, \aap, 483, 699

\bibitem[{{Hamuy} {et~al.}(2006){Hamuy}, {Folatelli}, {Morrell}, {Phillips},
  {Suntzeff}, {Persson}, {Roth}, {Gonzalez}, {Krzeminski}, {Contreras},
  {Freedman}, {Murphy}, {Madore}, {Wyatt}, {Maza}, {Filippenko}, {Li}, \&
  {Pinto}}]{hamuy06}
{Hamuy}, M., {et~al.} 2006, \pasp, 118, 2

\bibitem[{{Hennawi} {et~al.}(2009){Hennawi}, {Myers}, {Shen}, {Strauss},
  {Djorgovski}, {Fan}, {Glikman}, {Mahabal}, {Martin}, {Richards}, {Schneider},
  \& {Shankar}}]{hennawi09}
{Hennawi}, J.~F., {et~al.} 2009, ArXiv e-prints, 0908.3907

\bibitem[{{Hennawi} {et~al.}(2006){Hennawi}, {Strauss}, {Oguri}, {Inada},
  {Richards}, {Pindor}, {Schneider}, {Becker}, {Gregg}, {Hall}, {Johnston},
  {Fan}, {Burles}, {Schlegel}, {Gunn}, {Lupton}, {Bahcall}, {Brunner}, \&
  {Brinkmann}}]{hennawi06}
---. 2006, \aj, 131, 1

\bibitem[{{Hopkins} {et~al.}(2008){Hopkins}, {Hernquist}, {Cox}, \& {Kere{\v
  s}}}]{hopkins08}
{Hopkins}, P.~F., {Hernquist}, L., {Cox}, T.~J., \& {Kere{\v s}}, D. 2008,
  \apjs, 175, 356

\bibitem[{{Kauffmann} \& {Haehnelt}(2000)}]{kauffmann00}
{Kauffmann}, G., \& {Haehnelt}, M. 2000, \mnras, 311, 576

\bibitem[{{Kauffmann} {et~al.}(2003){Kauffmann}, {Heckman}, {Tremonti},
  {Brinchmann}, {Charlot}, {White}, {Ridgway}, {Brinkmann}, {Fukugita}, {Hall},
  {Ivezi{\'c}}, {Richards}, \& {Schneider}}]{kauffmann03}
{Kauffmann}, G., {et~al.} 2003, \mnras, 346, 1055

\bibitem[{{Kewley} {et~al.}(2001){Kewley}, {Dopita}, {Sutherland}, {Heisler},
  \& {Trevena}}]{kewley01}
{Kewley}, L.~J., {Dopita}, M.~A., {Sutherland}, R.~S., {Heisler}, C.~A., \&
  {Trevena}, J. 2001, \apj, 556, 121

\bibitem[{{Komossa} {et~al.}(2003){Komossa}, {Burwitz}, {Hasinger}, {Predehl},
  {Kaastra}, \& {Ikebe}}]{komossa03}
{Komossa}, S., {Burwitz}, V., {Hasinger}, G., {Predehl}, P., {Kaastra}, J.~S.,
  \& {Ikebe}, Y. 2003, \apjl, 582, L15

\bibitem[{{Kormendy} \& {Bender}(2009)}]{kormendy09}
{Kormendy}, J., \& {Bender}, R. 2009, \apjl, 691, L142

\bibitem[{{Kormendy} \& {Richstone}(1995)}]{kormendy95}
{Kormendy}, J., \& {Richstone}, D. 1995, \araa, 33, 581

\bibitem[{{Liu} {et~al.}(2010){Liu}, {Shen}, {Strauss}, \& {Greene}}]{liu10}
{Liu}, X., {Shen}, Y., {Strauss}, M.~A., \& {Greene}, J.~E. 2010, \apj, 708,
  427

\bibitem[{{Martini} {et~al.}(2004){Martini}, {Persson}, {Murphy}, {Birk},
  {Shectman}, {Gunnels}, \& {Koch}}]{martini04}
{Martini}, P., {Persson}, S.~E., {Murphy}, D.~C., {Birk}, C., {Shectman},
  S.~A., {Gunnels}, S.~M., \& {Koch}, E. 2004, in Society of Photo-Optical
  Instrumentation Engineers (SPIE) Conference Series, ed. {A.~F.~M.~Moorwood \&
  M.~Iye}, Vol. 5492, 1653--1660

\bibitem[{{Milosavljevi{\'c}} \& {Merritt}(2001)}]{milosavljevic01}
{Milosavljevi{\'c}}, M., \& {Merritt}, D. 2001, \apj, 563, 34

\bibitem[{{Moshir} {et~al.}(1992){Moshir}, {Kopman}, \& {Conrow}}]{moshir92}
{Moshir}, M., {Kopman}, G., \& {Conrow}, T.~A.~O. 1992, {IRAS Faint Source
  Survey, Explanatory supplement version 2}, ed. {Moshir, M., Kopman, G., \&
  Conrow, T.~A.~O.}

\bibitem[{{Myers} {et~al.}(2008){Myers}, {Richards}, {Brunner}, {Schneider},
  {Strand}, {Hall}, {Blomquist}, \& {York}}]{myers08}
{Myers}, A.~D., {Richards}, G.~T., {Brunner}, R.~J., {Schneider}, D.~P.,
  {Strand}, N.~E., {Hall}, P.~B., {Blomquist}, J.~A., \& {York}, D.~G. 2008,
  \apj, 678, 635

\bibitem[{{Owen} {et~al.}(1985){Owen}, {O'Dea}, {Inoue}, \& {Eilek}}]{owen85}
{Owen}, F.~N., {O'Dea}, C.~P., {Inoue}, M., \& {Eilek}, J.~A. 1985, \apjl, 294,
  L85

\bibitem[{{Rodriguez} {et~al.}(2006){Rodriguez}, {Taylor}, {Zavala}, {Peck},
  {Pollack}, \& {Romani}}]{rodriguez06}
{Rodriguez}, C., {Taylor}, G.~B., {Zavala}, R.~T., {Peck}, A.~B., {Pollack},
  L.~K., \& {Romani}, R.~W. 2006, \apj, 646, 49

\bibitem[{{Shen}(2009)}]{shen09}
{Shen}, Y. 2009, \apj, 704, 89

\bibitem[{{Skrutskie} {et~al.}(2006){Skrutskie}, {Cutri}, {Stiening},
  {Weinberg}, {Schneider}, {Carpenter}, {Beichman}, {Capps}, {Chester},
  {Elias}, {Huchra}, {Liebert}, {Lonsdale}, {Monet}, {Price}, {Seitzer},
  {Jarrett}, {Kirkpatrick}, {Gizis}, {Howard}, {Evans}, {Fowler}, {Fullmer},
  {Hurt}, {Light}, {Kopan}, {Marsh}, {McCallon}, {Tam}, {Van Dyk}, \&
  {Wheelock}}]{skrutskie06}
{Skrutskie}, M.~F., {et~al.} 2006, \aj, 131, 1163

\bibitem[{{Smith} {et~al.}(2009){Smith}, {Shields}, {Bonning}, {McMullen}, \&
  {Salviander}}]{smith09}
{Smith}, K.~L., {Shields}, G.~A., {Bonning}, E.~W., {McMullen}, C.~C., \&
  {Salviander}, S. 2009, ArXiv e-prints 0908.1998

\bibitem[{{Tody}(1986)}]{tody86}
{Tody}, D. 1986, in Society of Photo-Optical Instrumentation Engineers (SPIE)
  Conference Series, ed. {D.~L.~Crawford}, Vol. 627, 733

\bibitem[{{Toomre} \& {Toomre}(1972)}]{toomre72}
{Toomre}, A., \& {Toomre}, J. 1972, \apj, 178, 623

\bibitem[{{Veilleux} \& {Osterbrock}(1987)}]{veilleux87}
{Veilleux}, S., \& {Osterbrock}, D.~E. 1987, \apjs, 63, 295

\bibitem[{{Veilleux} {et~al.}(2001){Veilleux}, {Shopbell}, \&
  {Miller}}]{veilleux01}
{Veilleux}, S., {Shopbell}, P.~L., \& {Miller}, S.~T. 2001, \aj, 121, 198

\bibitem[{{Volonteri} {et~al.}(2003){Volonteri}, {Haardt}, \&
  {Madau}}]{volonteri03}
{Volonteri}, M., {Haardt}, F., \& {Madau}, P. 2003, \apj, 582, 559

\bibitem[{{Wang} {et~al.}(2009){Wang}, {Chen}, {Hu}, {Mao}, {Zhang}, \&
  {Bian}}]{wang09}
{Wang}, J., {Chen}, Y., {Hu}, C., {Mao}, W., {Zhang}, S., \& {Bian}, W. 2009,
  ArXiv e-prints

\bibitem[{{White} {et~al.}(1997){White}, {Becker}, {Helfand}, \&
  {Gregg}}]{white97}
{White}, R.~L., {Becker}, R.~H., {Helfand}, D.~J., \& {Gregg}, M.~D. 1997,
  \apj, 475, 479

\bibitem[{{Wyithe} \& {Loeb}(2003)}]{wyithe03}
{Wyithe}, J.~S.~B., \& {Loeb}, A. 2003, \apj, 595, 614

\bibitem[{{Xu} \& {Komossa}(2009)}]{xu09}
{Xu}, D., \& {Komossa}, S. 2009, \apjl, 705, L20

\bibitem[{{York} {et~al.}(2000){York}, {Adelman}, {Anderson}, {Anderson},
  {Annis}, {Bahcall}, {Bakken}, {Barkhouser}, {Bastian}, {Berman}, {Boroski},
  {Bracker}, {Briegel}, {Briggs}, {Brinkmann}, {Brunner}, {Burles}, {Carey},
  {Carr}, {Castander}, {Chen}, {Colestock}, {Connolly}, {Crocker}, {Csabai},
  {Czarapata}, {Davis}, {Doi}, {Dombeck}, {Eisenstein}, {Ellman}, {Elms},
  {Evans}, {Fan}, {Federwitz}, {Fiscelli}, {Friedman}, {Frieman}, {Fukugita},
  {Gillespie}, {Gunn}, {Gurbani}, {de Haas}, {Haldeman}, {Harris}, {Hayes},
  {Heckman}, {Hennessy}, {Hindsley}, {Holm}, {Holmgren}, {Huang}, {Hull},
  {Husby}, {Ichikawa}, {Ichikawa}, {Ivezi{\'c}}, {Kent}, {Kim}, {Kinney},
  {Klaene}, {Kleinman}, {Kleinman}, {Knapp}, {Korienek}, {Kron}, {Kunszt},
  {Lamb}, {Lee}, {Leger}, {Limmongkol}, {Lindenmeyer}, {Long}, {Loomis},
  {Loveday}, {Lucinio}, {Lupton}, {MacKinnon}, {Mannery}, {Mantsch}, {Margon},
  {McGehee}, {McKay}, {Meiksin}, {Merelli}, {Monet}, {Munn}, {Narayanan},
  {Nash}, {Neilsen}, {Neswold}, {Newberg}, {Nichol}, {Nicinski}, {Nonino},
  {Okada}, {Okamura}, {Ostriker}, {Owen}, {Pauls}, {Peoples}, {Peterson},
  {Petravick}, {Pier}, {Pope}, {Pordes}, {Prosapio}, {Rechenmacher}, {Quinn},
  {Richards}, {Richmond}, {Rivetta}, {Rockosi}, {Ruthmansdorfer}, {Sandford},
  {Schlegel}, {Schneider}, {Sekiguchi}, {Sergey}, {Shimasaku}, {Siegmund},
  {Smee}, {Smith}, {Snedden}, {Stone}, {Stoughton}, {Strauss}, {Stubbs},
  {SubbaRao}, {Szalay}, {Szapudi}, {Szokoly}, {Thakar}, {Tremonti}, {Tucker},
  {Uomoto}, {Vanden Berk}, {Vogeley}, {Waddell}, {Wang}, {Watanabe},
  {Weinberg}, {Yanny}, \& {Yasuda}}]{york00}
{York}, D.~G., {et~al.} 2000, \aj, 120, 1579

\end{thebibliography}

\end{document}